\documentclass[11pt]{article}

\usepackage{psfrag}
\usepackage{setspace}
\usepackage{graphics}
\usepackage{subfigure}
\usepackage{verbatim}
\usepackage{amssymb}
\usepackage{amsmath}
\usepackage{amsthm}
\usepackage{lscape}
\usepackage{epsfig}
\usepackage{color}
\usepackage{fullpage}
\usepackage{hyperref}

\newcommand{\beq}{\begin{equation}}
\newcommand{\eeq}{\end{equation}}
\newcommand{\beqr}{\begin{eqnarray}}
\newcommand{\eeqr}{\end{eqnarray}}
\newcommand{\beqrn}{\begin{eqnarray*}}
\newcommand{\eeqrn}{\end{eqnarray*}}
\newcommand{\beqn}{\begin{equation*}}
\newcommand{\eeqn}{\end{equation*}}
\newcommand{\bei}{\begin{itemize}}
\newcommand{\beii}{\begin{itemize} \item}
\newcommand{\eei}{\end{itemize}}
\newcommand{\bmei}{\begin{itemize} \compactlist}
\newcommand{\emei}{\end{itemize}}
\newcommand{\ben}{\begin{enumerate}}
\newcommand{\een}{\end{enumerate}}
\newcommand{\bes}{\begin{small}}
\newcommand{\ees}{\end{small}}
\newcommand{\bec}{\begin{center}}
\newcommand{\eec}{\end{center}}

\singlespace

\newcommand{\vect}[1]{\boldsymbol{#1}}
\newcommand{\naturalsWithZero}{\mathbb{N}_0}   
\renewcommand{\iff}{\mbox{$\Longleftrightarrow$}}    
\renewcommand{\implies}{\mbox{$\Longrightarrow$}}    

\begin{document}
\title{
\huge{\textbf{Impact of correlated neural activity on decision making performance}} \\[.5cm]
\Large{	Nicholas Cain and Eric Shea-Brown} \\
\Large{University of Washington, \\ 
Department of Applied Mathematics} \\ \vspace{1cm}
}\author{} \date{}
\maketitle
 \vspace{-.5in}

\begin{abstract}
	Stimulus from the environment that guides behavior and informs decisions is encoded in the firing rates of neural populations.  Each neuron in the populations, however, does not spike independently:  spike events are correlated from cell to cell.  To what degree does this apparent redundancy impact the accuracy with which decisions can be made, and the computations that are required to optimally decide?  We explore these questions for two illustrative models of correlation among cells.  Each model is statistically identical at the level of pairs cells, but differs in higher-order statistics that describe the simultaneous activity of larger cell groups.  We find that the presence of correlations can diminish the performance attained by an ideal decision maker to either a small or large extent, depending on the nature of the higher-order interactions.  Moreover, while this optimal performance can in some cases be obtained via the standard integration-to-bound operation, in others it requires a nonlinear computation on incoming spikes.  Overall, we conclude that a given level of pairwise correlations--even when restricted to identical neural populations--may not always indicate redundancies that diminish decision making performance.

\end{abstract}

%
%
%
%
%

\section{Introduction}

Sensory information is often encoded in irregularly spiking neural populations.  One well-studied example is 
given by direction-selective cells in area MT, whose firing rates depend on the degree and direction of coherent motion in the visual field~\cite{Britten:1993wv,Newsome:1989ul,Britten:1992wx,Salzman:1992wga}. 
Individual neurons in MT -- as in many other brain areas -- exhibit noisy and variable spiking \cite{Newsome:1989ul}, as can be modeled by Poisson point processes~\cite{Softky:1993uj,Tuckwell:1989us}.  Moreover, this variable spiking is generally not independent from cell to cell.  Returning to our example, a number of studies have measured pairwise correlations in MT during direction discrimination tasks as well as smooth-pursuit eye movements \cite{Huang:2009gc,Bair:2001wm,Zohary:1994uv,Cohen:2008js}; while this measurement is a subtle endeavor experimentally, a number of studies suggest a value near $\rho \approx .1-.15$ (\cite{Cohen:2011eh} summarizes these observations, for a number of brain areas.) 

What are the consequences of correlated spike variability for the speed and accuracy of sensory decisions?  The role of pairwise correlations in stimulus encoding has been the subject of many prior studies~\cite{Salinas:2001ul,2011arXiv1109.6524L,Averbeck:2006ew}.  The results are rich, showing that correlations can have positive, negative, or neutral effects on levels of encoded information.  The present study serves to extend this body of work in two ways.  First, as done in a different context by~\cite{Ganmor:2011ct,Montani:2009js}, we contrast the impact of correlations that have the same pairwise level but a different structure at higher orders.  

Second, as in \cite{Cohen:2008js,Beck:2008db}, we consider the impact of correlations on decisions that unfold over time, by combining a sequence of samples observed over time in the sensory populations.  A classical example that we will use to describe and motivate our studies is the {\it moving dots} direction discrimination task.  Here, a fraction of dots in a visual display move coherently in a given direction, while the remainder display random motion; the task is to identify the direction from two possible alternatives.  Decisions become increasingly accurate as subjects take (or are given) longer to make the decision.


In analyzing decisions that develop over time, we utilize a central result from sequential analysis.  This is the Sequential Probability Ratio Test (SPRT)~\cite{Wald:1948uha,Gold:2002th}, which linearly sums the log-odds of independent observations from a sampling distribution until a predetermined evidence threshold is reached.  The SPRT is the optimal statistical test in that it gives the minimum expected number of samples for a required level of accuracy in deciding among two task alternatives.  

We pose two related questions based on the SPRT.   
First, how does the presence of correlated spiking in the sampled pools impact the speed and accuracy of decisions produced by the SPRT?  Our focus is on how the structure of population-wide correlations determines the answer.  Second, how does the presence of correlated spiking impact the computations that are necessary to perform the SPRT?  This question is intriguing, because the SPRT may be performed via the simple, linear computation of integrating spikes over time and across the populations for a surprisingly broad class of inputs, including independent Poisson spike trains~\cite{Zhang:2010tj,Bogacz:2006fj}.  Thus, in this setting optimal decisions can be made by integrator circuits~\cite{Bogacz:2006fj,GoldmanReview,Cain:2012ky}.
Our goal here is to determine whether and when this continues to hold true for correlated neural populations.  



We answer these questions for two illustrative models of correlated, Poissonian spiking.  
We emphasize that the spikes that these models produce are indistinguishable at the level of both single cells and pairs of cells.  However, they differ in higher-order correlations, in that they can only be distinguished by examining the statistics of three or more neurons.  In the first model, 
correlations are introduced via shared spike events across the entire pool.  In this case optimal inference via the SPRT produces fast and accurate decisions, but depends on a nonlinear computation.  As a result, the simpler computation of spike integration requires, on average, longer times to reach the same level of accuracy. In contrast, when shared spiking events are more frequent but are common to fewer neurons within a pool, performance under the SPRT is significantly diminished.  However, in this case both SPRT and spike integration perform comparably, so a linear computation can produce decisions that are close to optimal.

\section{Models of evidence accumulation and encoding}

\subsection{Model neural populations and the decision task}

We begin by introducing the notation for the two decision making models that will be compared.  In this study we consider the case of discrimination between two alternatives, and therefore model two populations of neurons that encode the strength of evidence for each alternative.  Returning to the moving dots task for illustration, each population could be the set of MT cells that are selective for motion in a given direction.  Here, the firing rates in each population represents the dot motion $C$ via their firing rates $\lambda_p$ and $\lambda_n$; here the subscripts indicate the ``preferred" and ``null" populations, which correspond to the motion direction of the visual stimulus versus the alternate direction. In this way, the firing rate of neurons encoding the preferred direction will be higher than the null direction, $\lambda_p>\lambda_n$.  Following \cite{Wang:2002um} (see also \cite{Mazurek:2003cm,Britten:1993wv}), we model this relationship as linear:
\begin{eqnarray}
	\lambda_{p} &=& 40+.4C \; \text{Hz}\\
	\lambda_{n} &=& 40-.4C \; \text{Hz}.
\end{eqnarray}
Throughout the text we consider present results at $C=6.4$, however the results do not depend on this particular value of dot motion or its precise relationship firing rate.

In our model, we assume that each population consists of $N$ neurons firing spikes via a homogenous Poisson process, with rate $\lambda_p$ or $\lambda_n$.  We use the notation $x_k(t)$ to each spike train.  Integrating these processes over a time interval $\Delta T$ provides two time series of $N-$dimensional vectors of Poisson random variables; these independent vectors provide the input to the decision making models.  Specifically, for the $k^{\text{th}}$ neuron in a pool, on the $i^{\text{th}}$ time step,
\begin{equation}
	S^i_{k}
		= \int_{i\Delta T}^{(i+1)\Delta T}x_{k}(t)dt
		\sim \text{Poiss}(\lambda \Delta T).
\end{equation}
The properties of Poisson processes imply that $S^i_{k}$ is independent from $S^j_{k}$ ($i\neq j$), i.e. for different time steps.
 
However, the outputs of different neurons in the same time are not, in general, independent.  Following experimental observations that neurons with similar directional tuning tend to be correlated, while those with very different tuning are not~\cite{Zohary:1994uv,Cohen:2008js}, we model neurons from different pools as independent and those within a single pool as correlated with a correlation coefficient $\rho$:  
\begin{equation}
	\rho = \frac{\text{Cov}[S^i_{k},S^i_{l}]}{\sqrt{\text{Var}[S^i_{k}]\text{Var}[S^i_{l}]}},\;\; k\neq l.
					\label{eq:rho}	
\end{equation}
This implies that, with vector notation for the probability distribution of spike counts for each pool,
\begin{equation}
	P[\vect{S^i_p},\vect{S^i_n}] = P[\vect{S^i_p}]P[\vect{S^i_n}].
\end{equation}

Next, we introduce notation for decision making between the two task alternatives.  The task of determining, e.g., direction in the moving dots task is that of determining which of the two pools fires spikes with the higher firing rate.  We frame this as decision making between the hypotheses
			\begin{eqnarray}
				H_1 &:& \lambda_p>\lambda_n \\
				H_0 &:& \lambda_p<\lambda_n,
			\end{eqnarray}
where each alternative corresponds to a decision as to the motion direction. This formalism allows us to define accuracy as the fraction of trials on which the correct hypothesis $H_1$ is accepted.  In this study we consider decision making tasks at a fixed level of difficulty, so that $\lambda_p$ and $\lambda_n$ do not vary from trial to trial (i.e., this hypothesis test is simple and not composite).	 

\subsection{Accumulating spikes and evidence over time}

We relate the decision making task to a discrete random walk, which follows in turn from the sequential accumulation of independent and identically distributed (IID) realizations from the sampling distribution $W_i$.  We will specify this distribution below; for now, we note that the random walk takes the general form:
		\begin{equation}
			E_0 = 0
		\end{equation}
		\begin{equation}
			E_{n+1} = E_{n}+W_{n}, \;\; 
		\end{equation}
In a drift-diffusion model of decision making, accumulation continues as long as $|E_n| < \theta$, the decision threshold~\cite{Ratcliff:1978wz,Gold:2002th,Bogacz:2006fj}. The number of increments necessary to cross one of the two increments multiplied by its duration $\Delta T$ defines the decision time; this is a random variable, as it varies from trial to trial.  Crossing the threshold corresponding to $H_1$ is interpreted as a correct trial; the fraction of correct (FC) trials defines the accuracy of a the model. Together, the expected (mean) decision time ($DT$) and accuracy ($FC$) determine the performance of a decision making model.
		
	 Formulas for the mean decision time and accuracy are given in Wald \cite{Wald:1944th} as a function of the sampling distribution and the decision threshold.  Importantly, these formulas are exact under the assumption that the final increment in $E_n$ does not overshoot the threshold, a point we return to below. Given the moment generating function for the sampling distribution:
\begin{equation}
	\phi(s) = E[e^{W s}],
\end{equation}
Speed and accuracy are given by:
\begin{eqnarray}
	FC &\approx& \frac{1}{1+e^{h_0\theta}} \label{eq:FC} \\
	DT &\approx& \frac{\theta \Delta T}{E[W]} \tanh \left( \frac{-h_0 \theta }{2}\right) \label{eq:RT}
\end{eqnarray}
where $h_0$ is the nontrivial root of $\phi(s)-1$, i.e.
\begin{equation}
	\phi(h_0)=1,\;\;h_0\neq0.
\end{equation}
We notice here that as $\theta$ increases (and assuming $h_0<0$), both $FC$ and $RT$ will increase.
				
We now return to the definition of the random increments $W_i$.  We consider two different ways in which this can be done.  First, in the spike integration (SI) model, increments are constructed by counting the spikes emitted in a $\Delta T$ window by the preferred pool, and subtracting the number emitted by the null pool.  This is equivalent to the time evolution of a neural integrator model that receives spikes as impulses with opposite signs from the preferred and null populations.  This integrate-to-bound model is an analog of drift-diffusion model (DDM) with inputs that are not ``white noise", but rather Poisson spikes:
\begin{equation}
	W_i = \sum_{k=1}^N S^i_{k,p} - \sum_{k=1}^N S^i_{k,n}  \; \label{eq:SISamplingRV}
\end{equation}
\cite{Ratcliff:1978wz,Bogacz:2006fj,Zhang:2010tj,Beck:2008db}, cf.~\cite{Mazurek:2003cm}.

Second, in the Sequential Probability Ratio Test (SPRT), the increment is defined as the log-odds ratio of observing the spike count from both of the pools, under each of the two competing hypothesis:
\begin{equation}
	W_i = \log\left[	   \frac{P[\vect{S^i_p}|H_1]P[\vect{S^i_n}|H_1]}{P[\vect{S^i_p}|H_0]P[\vect{S^i_n}|H_0]}		\right] \label{eq:SPRTSamplingRV}
\end{equation}

\subsection{The case of independent neurons}

\cite{Zhang:2010tj} present an analysis of speed and accuracy of decision making based on independent neural pools; for completeness, and to help contrast this result with the correlated case, we give the key calculations in Appendices \ref{sec:MGF_SPRT},~\ref{sec:SIuncorr}.  Here, choosing increments via the SPRT yields:
	\begin{eqnarray}
		h_0 &=& -1  \label{eq:IndSPRTH0} \\
		E[W] &=& \Delta T N\left( \lambda_p-\lambda_n \right)\log \left( \frac{\lambda_p}{\lambda_n}\right). \label{eq:IndSPRTEz}
	\end{eqnarray}
	Under the spike integration model, Zhang and Bogacz \cite{Zhang:2010tj} (See also Appendix \ref{sec:SIuncorr}) find that:
	\begin{eqnarray}
		h_0 &=& -\log \left( \frac{\lambda_p}{\lambda_n}\right). \label{eq:uncorrSI}\\
		E[W] &=& \Delta T N\left( \lambda_p-\lambda_n \right).\label{eq:uncorrSIEz}
	\end{eqnarray}	
	 Therefore, by applying a change of variables $\theta \rightarrow \theta \log \left( \frac{\lambda_p}{\lambda_n} \right)$ in Equations ~\ref{eq:FC} and \ref{eq:RT}, spike integration can implement the SPRT. The implication is that simply counting spikes, positive for one pool and negative for the other, can implement  statistically optimal decisions for when the neural pools are independent \cite{Zhang:2010tj}. 
	 
\subsection{Correlated neural populations: the additive and subtractive models} \label{s.corr}

	 We next describe two models for introducing correlations into the Poisson spike trains of each neural population.
	 Both models are studied in \cite{Kuhn:2003tq,springerlink:10.1007/978-1-4419-5675-0_12}, and rely on shared input from a single correlating process to generate the correlations in each pool.  These authors termed the two model SIP and MIP for single- and multiple-interaction process; here we use the added descriptors ``additive" and ``subtractive." In both models, a realization of correlated spike trains that provide the input to the accumulation models is achieved via a common correlating train.
	 
	 Before describing the models in detail, we note that in this study, these models are statistical approaches chosen to illustrate a range of impacts that correlations can have on decision making (see also~\cite{Gutnisky:2010cp,Niebur:2007to}).  In contrast, in neurobiological networks, correlated spiking arises as through a complex interplay of many mechanisms, including recurrent connectivity and shared feedforward interactions (For example,~\cite{Aertsen:1989tg, Shadlen:1998ta, Smith:2008gv}).  
While beyond the scope of the present paper, avenues for bridging the gap between statistical and network-based models of correlations in the context of decision making are considered in the Discussion.
	 
\begin{figure}
	\centering
	\includegraphics[width=6in]{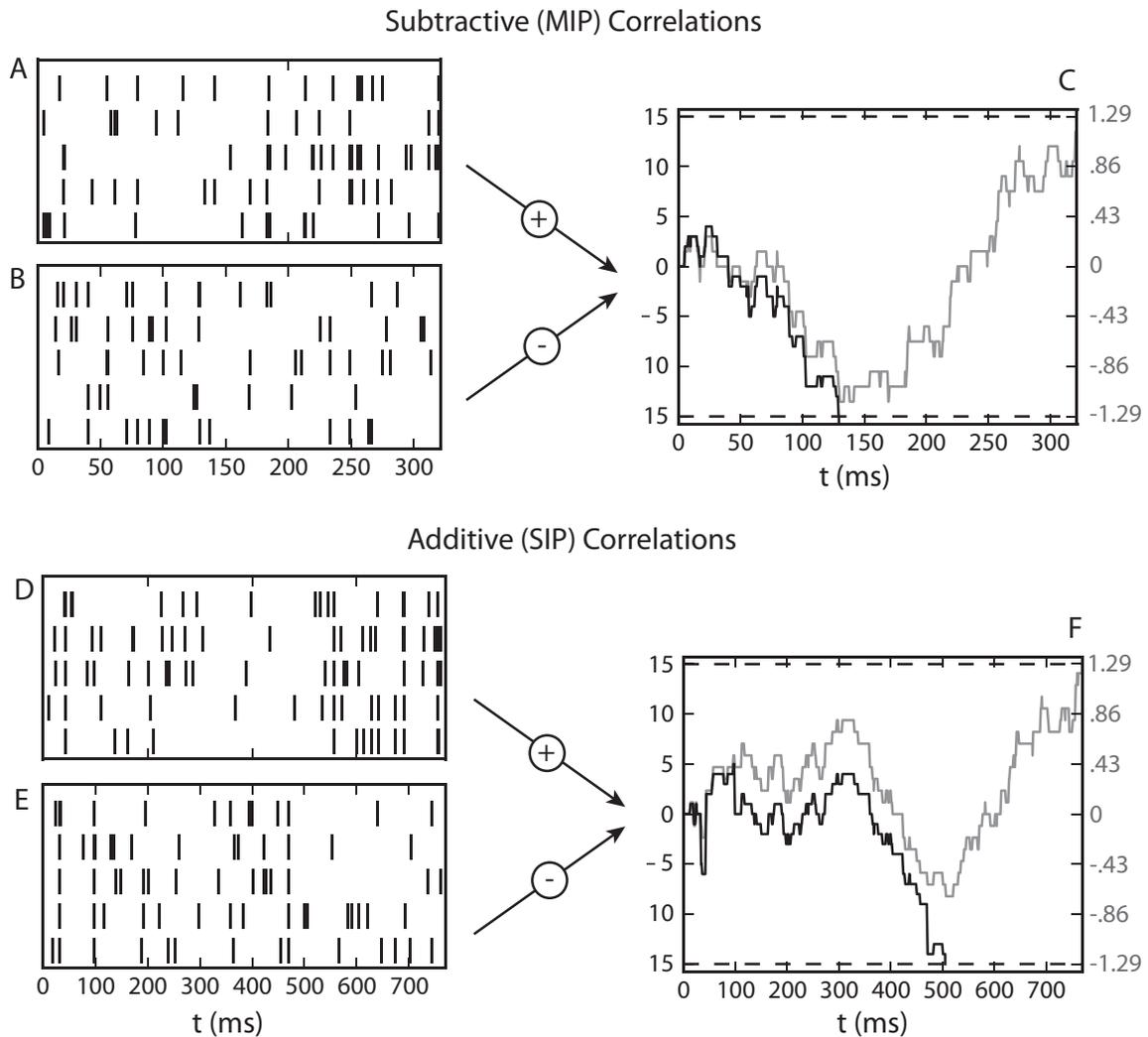}
	\caption{Spike integration (SI) and SPRT for a single trial, with subtractive (MIP) correlations (A,B,C) and additive (SIP) correlations (D,E,F). Rastergrams at C$=6.4$ for preferred (A,D) and null (B,E) populations of $5$ neurons, with spike count correlation within pools $\rho=.15$. In (C,F) these spikes are either integrated (black line) or provide input for the SPRT (gray line), until a decision threshold is reached. The decision threshold has been set so that all four cases will yield the same mean reaction time (In C, $\theta_{SI}=15$ and $\theta_{SPRT}=1.28$, and in F $\theta_{SI}=14$ and $\theta_{SPRT}=1.28$; in both cases the SPRT lines have been scaled for plotting purposes).  On these trials, the SPRT accumulator crosses the ``correct", upper, threshold, as opposed to the ``incorrect", lower, threshold for the spike integrator.  Unlike the independent case, the time evolution of the spike integration process is not simply a scaled version of the SPRT (though they are clearly similar) under either model of correlations.}
	\label{fig:IntSPRTOverview}
\end{figure}

The first case is the additive (SIP) model, in which the spike train for each neuron is generated as the sum of two homogenous Poisson point processes. 
The first Poisson train is generated with an overall firing rate of $(1-\rho) \lambda$, where $\lambda$ is the intended firing rate of the neuron, and $\rho$ is the intended pairwise spike count correlation between any two neurons in the pool.  The second train, with a rate of $\rho \lambda$, is added to every neuron in the pool, and serves as the common source of correlations. An example of this model of spike train generation is depicted in the rastergrams in Figure \ref{fig:IntSPRTOverview}A and B; the common spike events are evident as shared spikes across the entire population. 

The second case is the subtractive (MIP) model, in which correlated spikes are generated through random, independent deletions from an original ``mother" spike-train; we refer to this as the correlating spike train~\cite{Kuhn:2003tq}.  There is a separate correlating spike train for each of the two independent populations.  In order to achieve an overall firing rate for the pool of $\lambda$ spikes per second, with a pairwise correlation $\rho$ between any two individual neurons, the correlating train has a rate of $\lambda / \rho$ spikes per second.  Then, for each neuron in the pool, a spike is included from this train IID with a probability of $\rho$. An example of this model of spike train generation is depicted in the rastergrams in Figure \ref{fig:IntSPRTOverview}D and E. 

In summary, the two models both include correlated spike events that originate in from a single ``mother." Although they produce identical correlations among cell pairs, these events are distributed in different ways across the entire population.  
We note that the results of \cite{Zhang:2010tj} can be seen as a limiting case as $\rho\rightarrow0$ of either the additive (SIP) or subtractive (MIP) models.  
	 
\section{Subtractive (MIP) correlations and decision making performance}
\subsection{The SPRT decision making model}

We now study the impact of subtractive (MIP) correlations on decision making performance.
As noted above, recall that within a time window $\Delta T$, the spike counts from each neuron form a vector of random variables which are independent from window to window. These independent vectors provide the evidence for each of the two alternatives, which is then weighed via log-likelihood at each step in SPRT.  In Appendices \ref{sec:MGF_SPRT} and  \ref{sec:MIPSPRTAppendix}, we compute the values $h_0$ and $E[W]$ that define the speed and accuracy of the SPRT (see Equations~\ref{eq:FC}-\ref{eq:RT}), for two pools with subtractive (MIP) correlations.  As this computation is done in continuous time, it is natural to take $\Delta T \rightarrow 0$; doing so, we find:

\begin{equation}
	h_0=-1
\end{equation}
\begin{equation}
	E[W] =  \frac{1-(1-\rho)^N}{\rho}(\lambda_p-\lambda_n)\log \frac{\lambda_p} {\lambda_n} \Delta T+O(\Delta T^2) \label{eq:MIPSPRTEz}
\end{equation}

Comparing these values against those of the independent SPRT given in Equations \ref{eq:IndSPRTH0} and \ref{eq:IndSPRTEz}, we see that the only effect of correlations is a scaling of the expected increment via $\left(1-(1-\rho)^N \right)/\rho$.  In the limit as $\rho\rightarrow0$, this scale factor approaches N, which in turn reduces decision time (the scale factor is inversely proportional to $DT$ via Equation \ref{eq:RT}).  On the other hand, as $\rho\rightarrow1$, the scale factor itself approaches 1; this agrees with the intuition that as all neurons become perfectly redundant, the performance should resemble that of a single neurons. In fact, the mechanism of the SPRT on a given sample can be seen as inferring the firing rate of the correlating train from a derived vector of noisy random variables.  As $N$ gets large, then, performance should be limited by performing an SPRT on the correlating ``mother" trains themselves.  This is precisely what happens when $N\rightarrow\infty$ in Equation \ref{eq:MIPSPRTEz}:  we obtain $E[W] \sim \frac{1}{\rho}(\lambda_p-\lambda_n)\log \frac{\lambda_p} {\lambda_n} \Delta T,$ corresponding to decision making based on mother spikes of rate ${\lambda_p}/\rho$ and ${\lambda_n}/\rho$.

One consequence of this interpretation is that the particular realization of a spike vector (in a sufficiently small time-bin $\Delta T$) carries no evidence about the decision of $H_1$ vs. $H_0$, beyond its identity as either the zero vector $\vect{0}$ or not.  Of course, this is a consequence of the construction of the MIP model, as the spike deletions that create the realization of the spike vector have no dependence on the firing rate of the population. Concretely then, the increments (or decrements) are based solely on whether the vector of spikes in the preferred (or null) pool contains any spikes at all; the actual number of spikes is irrelevant in the SPRT.

It follows that the accumulation process $E_n$ is a discrete-space random walk, with steps $\pm\log(\lambda_p/\lambda_n)$.  To see this, note that for sufficiently small $\Delta T$, there are only three possibilities for how spikes will be emitted from the two populations.  First, both the preferred and null pools could produce no spikes.  This event provides no information to distinguish the firing rates of the pools, so the increment is 0.  Second, one of the pools could produce a vector of spikes caused by IID deletions from the ``mother" spike train.  If the spiking pool is the preferred one, each possible nonzero spike vector will increment the accumulator by the $\log$ of the ratio $(\lambda_p/\rho)/(\lambda_n/\rho)$; the opposite sign occurs if the null pool spikes.  Events in which both pools spike are of higher order in $\Delta T$, and thus become negligible for small time windows.

\begin{figure}
	\centering
	\includegraphics[width=6in]{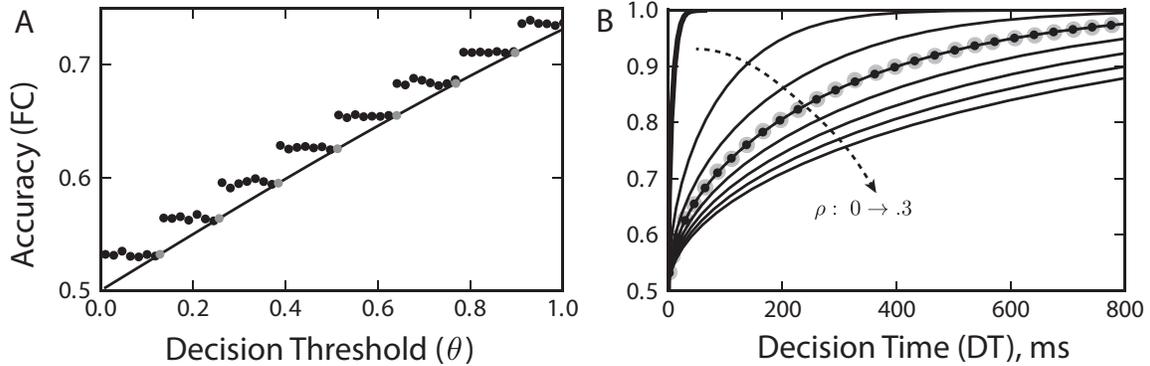}
	\caption{Subtractive (MIP) correlations significantly diminish decision performance under SPRT (C$=6.4$, $N=240$). (A) The discrete nature of the SPRT diffusion process implies that only discrete values of accuracy are possible.  These occur at values of $\theta$ that are multiples of $\log(\lambda_p/\lambda_n)\approx.1.28$. (Similar results hold for decision time, not shown.) The solid dots are simulations of the SPRT, and gray dots are exact values taken at multiples of the log ratio; the interpolating line is Equation \ref{eq:FC}. (B) Accuracy (Equation \ref{eq:FC}) and decision time (Equation \ref{eq:RT}) are plotted parametrically as a function of threshold, for 8 different values of $\rho$ (linearly spaced on [0,.35] with the a double-thickness line at $\rho=0$). Performance of the simulation at multiples of the log-ratio of firing rates are plotted as solid dots, and theoretical values in gray (gray dots are enlarged to be distinguished).}
	\label{fig:MIPSPRT}
\end{figure}

The discrete nature of the SPRT effect causes the $FC$ curve in Figure \ref{fig:MIPSPRT}(A) to take on only discrete values of accuracy; a small increase in $\theta$ above a multiple of $\log(\lambda_p/\lambda_n)$ will not improve accuracy because $E_n$ on the final, threshold-crossing-step will overshoot the threshold. This also explains why some of the $FC$ values at a given $\theta$ do not lie on the theoretical line defined by Equation $\ref{eq:FC}$; that equation is only exactly true in the case of zero overshoot past the threshold.  We will return to this point later, and also in Appendix \ref{sec:overshootAppendix}.

We next insert the values for $h_0$ and $E[W]$ computed above into Equations \ref{eq:FC} and \ref{eq:RT}, and plot the resulting speed-accuracy curves relating $DT$ and $FC$ parametrically in the threshold $\theta$ (Figure \ref{fig:MIPSPRT}(B)).  (We plot the full $FC$ and $RT$ functions, although only discrete values of performance along each of the lines are achievable in practice, as indicated by the dots for the $\rho=.15$ case; see caption).  By comparing speed-accuracy curves for different values of $\rho$ ranging from 0 to 0.3, we see our first main result:  {\it introducing MIP correlations within neural populations substantially diminishes the best-possible decision performance, that obtained via the SPRT.} We will next derive the analogous results for the simpler spike integration model.

\subsection{The spike integration decision making model}

\begin{figure}
	\centering
	\includegraphics[width=3in]{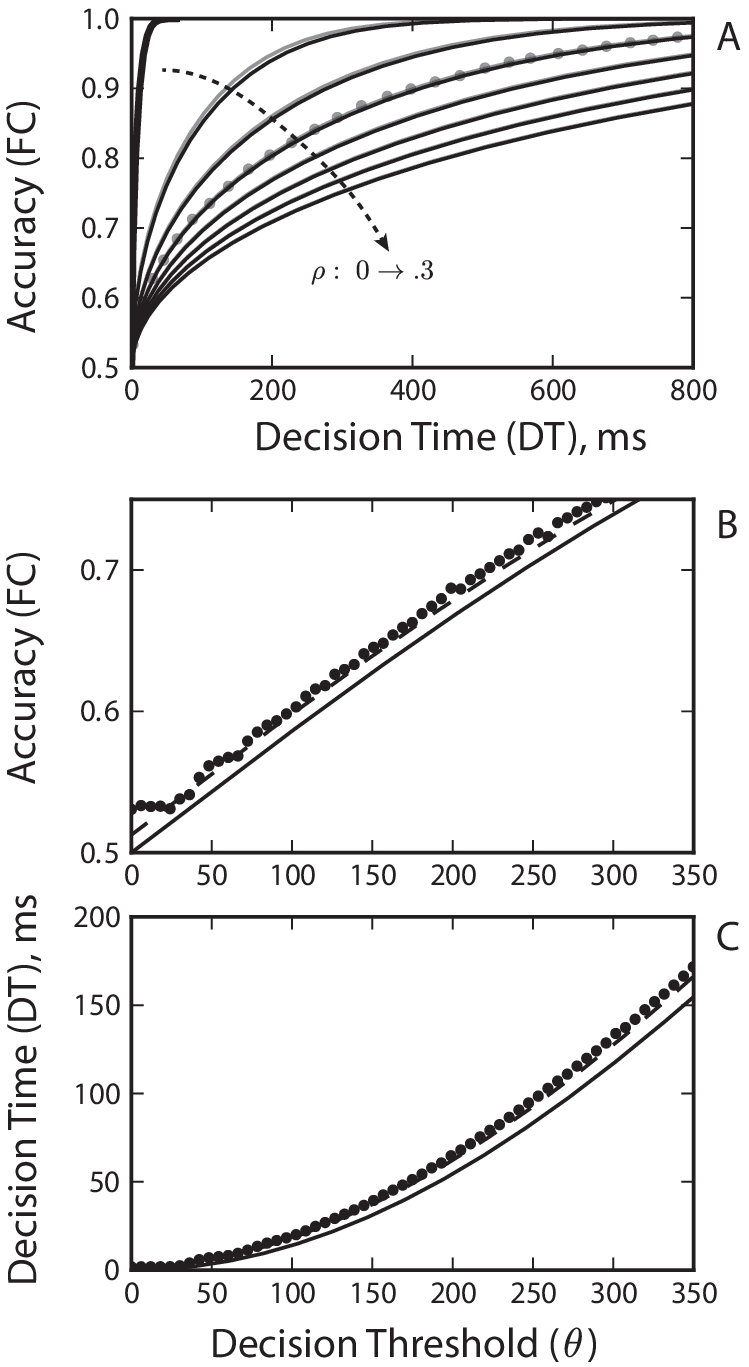}
	\label{fig:MIPSI}
	\caption{For the subtractive (MIP) model of spike correlations, decision making performance of the spike integration model is comparable to the SPRT, and is well described by Equations \ref{eq:FC} and \ref{eq:RT} despite overshoot past past the decision threshold. (A) Gray lines are reproductions of speed accuracy curves from the SPRT (Figure \ref{fig:MIPSPRT}), and black lines are speed accuracy curves for spike integration. (B,C) Overshoot past the decision boundaries reduces the validity of Wald's approximations, but a constant shift in threshold can help mitigate the effect (See \cite{Ghosh:1991wm, Lee:1994um} and Appendix \ref{sec:overshootAppendix}).  Such a shift is automatically accounted for when comparing curves that are parametric in $\theta$ (Panel A, for example).}
\end{figure}

Next, we consider decision making performance for the simpler model in which spikes are simply integrated over time, as opposed to the likelihood ratio computation of the SPRT.  In this case, the moment generating function of the difference in spike counts from the two pools is more straightforward (See Appendix \ref{sec:MIPSIAppendix}), and provides an easy computation of $E[W]$:  
\begin{equation}
	E[W]=\Delta T N\left( \lambda_p-\lambda_n \right) \label{eq:MIPSIimplicitEz}
\end{equation}
The nontrivial root of the MGF $h_0$ is found to be the implicit solution of:
\begin{equation}
	\left( 
			   ((1+\rho(e^t-1))^N -1)
	  \right)\lambda_p + 
	\left( 
			   ((1+\rho(e^{-t}-1))^N -1)
	  \right)\lambda_n=0 \label{eq:MIPSIimplicith0}
\end{equation}
Here we see that correlations only impact the performance of the model through changing $h_0$, as the expected increment is the same is in the independent case (Equation \ref{eq:uncorrSIEz}). 
Moreover, performance under spike integration is diminished to a degree that is comparable to the performance loss of SPRT.  To illustrate this, Figure \ref{fig:MIPSI}A plots the speed-accuracy tradeoff curves from both models of decision making under subtractive correlations, for the same values of $\rho$.  As we must(~\cite{Wald:1948uha}), we see the optimal character of the SPRT in the fact that at a given level of accuracy, the SPRT requires, on average, fewer samples than spike integration.  However, the difference is very slight.  This yields our next main result, that {\it nearly optimal decisions are produced by the simple operation of linear integration over time for the MIP model of spike correlations across neural populations.}


Having established this, we pause to note a subtlety in our analysis.  Figures \ref{fig:MIPSI}B and C show $FC$ and $DT$ as a function $\theta$, for both simulated data and plots of Equations $\ref{eq:FC}$ and \ref{eq:RT}.  The solid lines are the graphs of those equations as written (using the values for $h_0$ and $E[W]$ in Equations \ref{eq:MIPSIimplicith0} and \ref{eq:MIPSIimplicitEz}), and the mismatch between the lines and the data are a consequence of overshoot past the threshold.  The broken line is a graph of the same formulas, with a shift in $\theta\rightarrow\theta+14.5$, an offset computed as the sample mean of the overshoot distribution (See Figure \ref{fig:OvershootMIP} as well as the discussion in Appendix \ref{sec:overshootAppendix}; also \cite{Ghosh:1991wm, Lee:1994um}).  This correction term helps the $FC$ and $DT$ equations better approximate the data when there is potential overshoot. Interestingly, however, parametric plots like Figure \ref{fig:MIPSI}A already take this effect into account.


\section{Additive (SIP) correlations and decision making performance}
\subsection{The SPRT decision making model}

As described in Section \ref{s.corr}, the additive (SIP) model of spike train correlations also utilizes a common spike train to generate correlations, but does so in a manner that gives a distinct population-wide correlation structure.  We now derive the consequences for decision making performance under the SPRT.  In Appendices \ref{sec:MGF_SPRT} and \ref{sec:SIPSPRTEwAppendix} we find the expressions for the parameters of the $FC$ and $DT$ curves, as the window size $\Delta T\rightarrow0$:
\begin{equation}
	h_0=-1
\end{equation}
\begin{equation}
	E[W] = \left(N(1-\rho)+\rho\right)  (\lambda_p-\lambda_n) \log \frac{\lambda_p} {\lambda_n}\Delta T+O(\Delta T^2) \label{eq:SIPSPRTEz}
\end{equation}
Comparing these with Equations \ref{eq:IndSPRTH0} and \ref{eq:IndSPRTEz}, we see that, as in the subtractive (MIP) correlations model, the only difference with the independent case is a scaling factor on the average increment $E[W]$ in Equation \ref{eq:SIPSPRTEz}.  
To explain the form of the scale factor, note that the spike vector from each pool is composed of $N$ independent spike trains firing at rate $\lambda(1-\rho)$, and a single (highly redundant) spike train firing at a rate $\lambda \rho$.  

\begin{figure}
	\centering
	\includegraphics[width=6in]{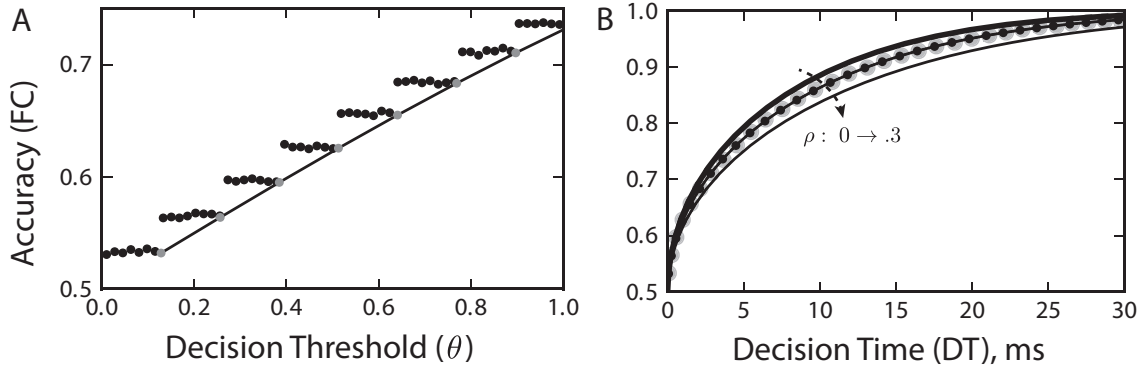}
	\caption{Additive (SIP) correlations do {\it not} significantly diminish decision performance under the SPRT. (A) The discrete diffusion with increment $\pm\log(\lambda_p/\lambda_n)\approx.128$ gives the same accuracy as the subtractive (MIP) correlations case (Figure \ref{fig:MIPSPRT}A) {\it at each value of $\theta$.} Because of the absence of overshoot, the $FC$ and $DT$ relationships can be applied exactly. (B) However, the resulting speed-accuracy curves are very different.  In particular the impact of correlations on the speed-accuracy tradeoff is much smaller than for subtractive correlations (cf. Figure \ref{fig:MIPSPRT}B, noting that here the abscissa ranges up to 30 ms, in contrast to 800 ms).  Here only $\rho=0, 0.15,$ and $0.3$ are plotted for clarity.}
	\label{fig:SIPSPRT}
\end{figure}

As in the subtractive (MIP) model, $E_n$ here also becomes a discrete random walk with increment $\pm \log (\lambda_p/\lambda_n)$.  This can be seen by noting that for either pool, in a sufficiently small $\Delta T$ window, only one of two events is possible: (i) no spikes occur at all, or (ii) a single spike occurs in one neuron, in one of the two pools.
The first case is uninformative about either $H_1$ or $H_0$. The second case occurs with probability $\lambda(1-\rho)$ under $H_1$ and $\lambda(1-\rho)$ under $H_0$ (Here $\lambda=\lambda_p$ if the spike occurred in the preferred pool, for example); taking the log ratio, we find our increment is independent of correlations.
The resulting decision accuracy (FC) is plotted vs. threshold in Figure \ref{fig:SIPSPRT}A, and is qualitatively similar to the subtractive (MIP) correlations case, with plateaus following from the discrete nature of $E_n$. However, the speed-accuracy tradeoff pictured in Figure \ref{fig:SIPSPRT}B is very different from that found in the subtractive (MIP) model.  

In particular, we see our third main result:  {\it the impact of additive correlations on optimal (SPRT) decision performance is relatively minor.}  For example, in the presence of pairwise correlations as strong as $\rho=.3$, the mean decision time required to reach a typical value of accuracy is increased by only a few milliseconds compared with the independent case, instead of by hundreds of milliseconds as for subtractive correlations.  Equation \ref{eq:SIPSPRTEz} offers an intuitive explanation for this fact: $E[W]$ is inversely proportional to $DT$, and does not diminish nearly as fast for SIP correlations than MIP correlations (cf. Equation \ref{eq:MIPSPRTEz}).

\subsection{The spike integration decision making model}

\begin{figure}
	\centering
	\includegraphics[width=6in]{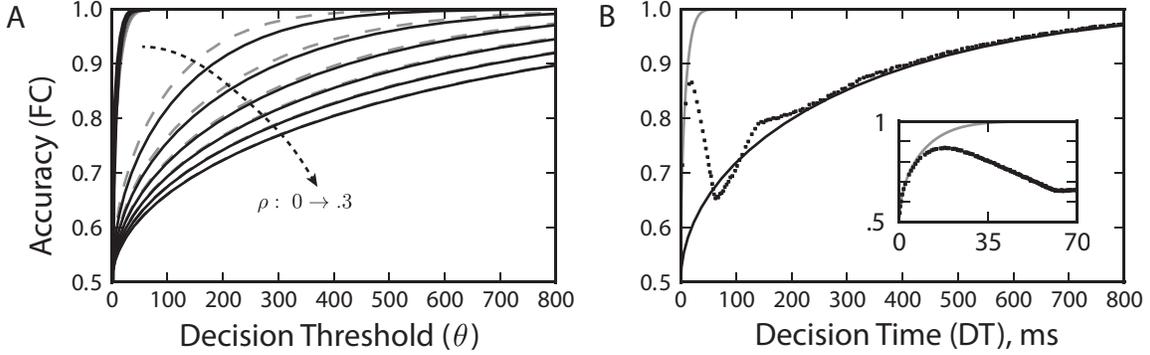}
	\caption{Decision making performance of the spiking integrator model with additive (SIP) correlations is comparable to subtractive correlations:  correlations significantly decrease performance. (A) Black lines give the speed accuracy tradeoff predicted using $h_0$ and $E[W]$ from Equations \ref{eq:SIPSIh0MainText} and \ref{eq:SIPSIEzMainText} (and thereby assuming no overshoot of the decision threshold).  Performance is similar to the subtractive-correlations case (broken gray lines), and significantly worse than performing SPRT on additive-correlated inputs (solid gray lines). (B) At $\rho=.15$, for example, major differences arise between this theory (again, solid black line, reproduced from A) and simulation of the model (dots), especially at short reaction times. This is a consequence of significant overshoot of $E_n$ over the decision threshold, on the threshold crossing step. (Inset) At short reaction times, the simulations actually perform closer to the SPRT (gray line, reproduced from Figure \ref{fig:SIPSPRT}A); see text.}
	\label{fig:SIPSI}
\end{figure}

What about the ability of the simple spike integrator to perform decision making when confronted with additive correlations? Proceeding as in the subtractive-correlations case, we derive an implicit relationship for $h_0$, and the expected increment $E[W]$:
\begin{equation}
	\lambda_p(
	\rho(e^{Nt}-1)+(1-\rho)N(e^t-1)
			 ) 
	+
	\lambda_n(
	\rho(e^{-Nt}-1)+(1-\rho)N(e^{-t}-1)
			 ) = 0 \; \iff \; h_0=t \label{eq:SIPSIh0MainText}
\end{equation}
\begin{equation}
	E[W]=\Delta T N\left( \lambda_p-\lambda_n \right) \label{eq:SIPSIEzMainText}
\end{equation}
By comparing with (Equation \ref{eq:uncorrSIEz}), we see that, as for spike integration in the subtractive (MIP) case, correlation affects only the value of $h_0$ and not the expected increment.  Substituting these values into Equations \ref{eq:FC} and \ref{eq:RT}, we then plot the speed-accuracy tradeoff curves for this model {\it under the assumption of no overshoot} in Figure \ref{fig:SIPSI}A. It appears that, when decisions are made via spike integration, correlations impact performance quite significantly (black lines), in contrast to the SPRT case (solid gray lines, reproduced from Figure \ref{fig:SIPSPRT}B).  Overall, the degree of performance loss is comparable to that under subtractive correlations (broken gray lines, reproduced from Figure \ref{fig:MIPSI}B).  This is our fourth main result:  {\it for additive correlations, if decisions are made via spike integration instead of the SPRT, correlations have a significant impact on reducing decision performance.}

However, the assumption that integrated spikes do not overshoot the decision threshold might seem suspect under the additive model of correlations, as there is a possibility that the threshold crossing step might occur as a result of every neuron in a pool simultaneously spiking at once.   In fact, when the number of neurons in the pool is large (as in the cases we consider), additive correlations can indeed cause significant overshooting of thresholds; importantly, and unlike for subtractive (MIP) correlations, this effect cannot be compensated via a constant offset in the decision threshold.  

Figure \ref{fig:SIPSI}B demonstrates the consequences for the speed-accuracy tradeoff.  Here, when the spike integration model is simulated directly, we see a surprising non-monotonic relationship between $FC$ and $DT$ in the presence of additive correlations of strength $\rho=.15$.  This violates the usual intuition of that accuracy should increase at slower decision speeds.  The explanation comes from the fact that, as the decision threshold is raised increases, $DT$ correspondingly increases while accuracy suffers -- a consequence of not finishing a trial before a (relatively rare) spike in a correlating spike train in one of the two pools causes the accumulator to jump far beyond the threshold.

\begin{figure}
	\centering
	\includegraphics[width=6in]{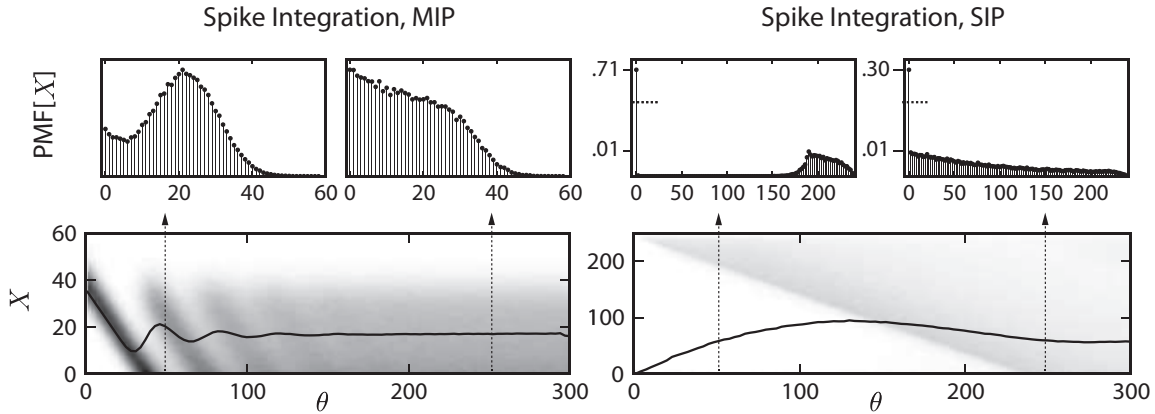}
	\caption{Overshoot distributions for spike integration under additive (SIP) and subtractive (MIP) correlations.  The random variable $X$ indicates the distribution of $E_n-\theta$ conditioned on crossing the upper threshold (similar results for the lower threshold are not shown).  The probability mass function (PMF) of $X$ varies as a function of $\theta$, and two vertical slices through this density are shown at $\theta=50$ and $250$.  Here the overshoot distributions are discrete, due to the integral nature of the increment distribution.  For plotting purposes, the vertical axis has been split in the SIP case, to allow plotting of the outlier point at zero.  The black line indicates $E[X]$ as $\theta$ varies; crucially, this quantity varies significantly and for higher values of $\theta$ under SIP correlations, resulting in the non-monotonic speed-accuracy tradeoff pictured in Figure \ref{fig:SIPSI}.}
	\label{fig:OvershootMIP}
\end{figure}

For large thresholds, the sequential sampling theory of Equations \ref{eq:FC} and \ref{eq:RT}, which assume no overshoot, accurately approximates the simulated data; however for low values of $\theta$ the approximation is poor. In fact, the inset to Figure \ref{fig:SIPSI}B shows that in this regime, the decision making performance of the spike integration model is far better described by the theory predicted by the SPRT.  The intuition behind this observation is that for short reaction times, there is a small probability of a shared spike that will send the integrator significantly over the threshold.  This allows accumulation to occur one spike at a time (for sufficiently small $\Delta T$), where each spike arrives from an independent spike train.  As we have seen, the process of integrating independent spikes is equivalent to the SPRT.  It is only at longer decision times, when the chances of having integrated a large common spike event are larger, that a significant impact of correlations appears.  

Figure \ref{fig:Nonlinearity} provides further evidence for this scenario.  Density plots of the distribution of the overshoot $X=E_x-\theta$ (conditioned on crossing the upper threshold), for both additive (SIP) and subtractive (MIP) correlations are shown as a function of the decision threshold, with particular overshoot distributions plotted at $\theta=50$ and $250$.  For the additive correlations model, a significant fraction of the trials terminate with zero overshoot at low values of $\theta$ (because, for example, large correlating events are relatively rare), implying that many trials underwent optimal accumulation of evidence, without experiencing a common, correlating spike event as discussed above. 

Overall, the monotonic dependence of accuracy (FC) on decision time ($DT$) follows from the invariance of the moments of the overshoot distribution relative to changes in the threshold value $\theta$; this is particularly true for the first moment (See Appendix \ref{sec:overshootAppendix}).  Figure \ref{fig:Nonlinearity}(SIP) demonstrates that these moments continue to fluctuate over a larger range of $\theta$, and with larger magnitude, for the additive correlations model.  This serves to explain the strange shape of the speed-accuracy tradeoff curve pictured in Figure \ref{fig:SIPSI}B that (unlike the subtractive correlations model) cannot be explained by a constant shift in $\theta$.

\section{Nonlinear computations and optimal performance via the SPRT}

\begin{figure}
	\centering
	\includegraphics[width=6in]{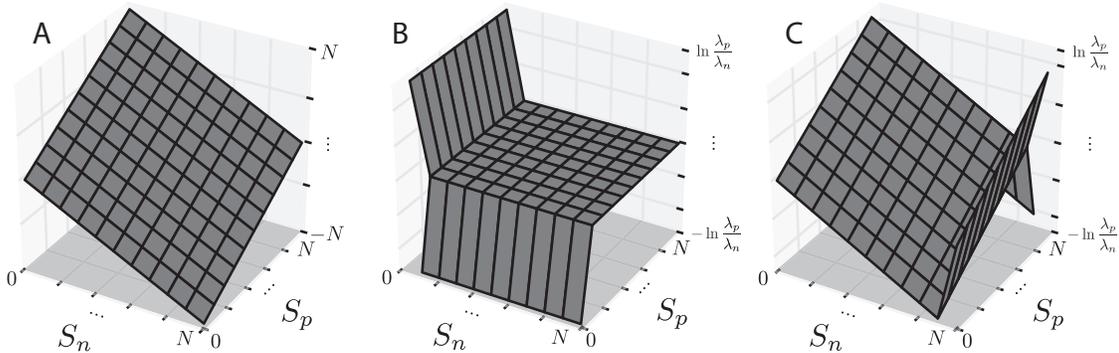}
	\caption{Increments for the SPRT are nonlinear when input spikes are correlated.  (A) For both additive and subtractive correlations, the spike integration model of decision making implies a linear mapping between the number of spikes in the preferred and null populations, and the increment to the accumulator. (B) With subtractive correlations, a severe nonlinearity means that only increments of $\pm \log (\lambda_p/\lambda_n)$ occur. This stands in direct contrast to the optimality of linear summation in the zero-correlations case.  (C) A nonlinear computation also appears as a consequence of the additive correlations model, however the nonlinearity is much less severe than in the subtractive model. (All results pictured hold in the case of vanishing $\Delta T$.)}
	\label{fig:Nonlinearity}
\end{figure}

\begin{figure}
	\centering
	\includegraphics[width=6in]{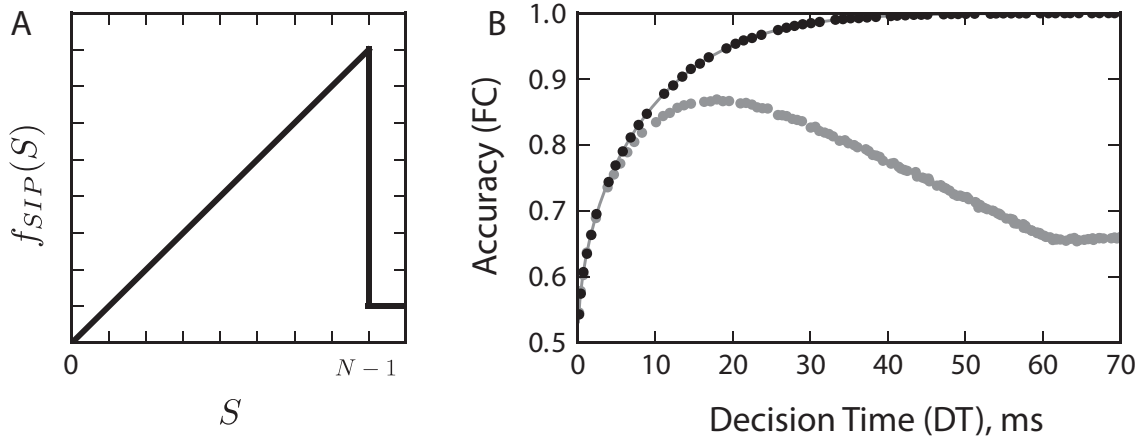}
	\caption{Optimal performance via spike integration under additive correlations can be realized with a simple nonlinearity. (A) A nonlinearity discounts the contribution to the accumulator of a shared spike event (See Equation \ref{eq:transferFunction}). (B) Spike integration with this nonlinearity is suggested by Figure \ref{fig:Nonlinearity}C, and recovers performance of the decision making model (black dots) to  agreement with the results of SPRT (gray line). Without this nonlinearity to discount shared events, performance suffers (gray dots, reproduced from Figure \ref{fig:SIPSI}B, Inset)}
	\label{fig:SIPSINonlinear}
\end{figure}

When the neurons in each pool spike independently, Zhang and Bogacz \cite{Zhang:2010tj} demonstrated that linear summation of spikes across the two pools at each time step implements the SPRT.  Because the SPRT is optimal in the sense of minimizing $DT$ for a prescribed level of $FC$, the conclusion is that linear integration of spikes across pools, and then across time, provides an optimal decision making strategy. However, is this optimality of linear integration confined to the case of independent activity within the pool?

Above, we showed that when correlations are introduced into this model, it is no longer true that each spike should be given the same ``weight", as in linear integration. Moreover, knowing only the pairwise correlations and firing rates alone does not allow one to write down a rule for the function that should be applied to incoming spikes in order to implement the SPRT, although in these cases this function takes the form of the difference between the result of a nonlinearity applied to both pools.  This dependence on higher order statistics is demonstrated in Figure \ref{fig:Nonlinearity} by the fact that the nonlinearities for MIP correlations (Panel B) and SIP correlations (Panel C) take a significantly different form.

For MIP correlations, the nonlinearity pictured in Figure \ref{fig:Nonlinearity}B that implements the SPRT (up to a change in threshold) takes the form:
\begin{equation}
	W_i = f_{MIP}(\vect{S}^i_p)-f_{MIP}(\vect{S}^i_n) \label{eq:transferFunctionMIP}
\end{equation}
\begin{equation}
	f_{MIP}(\vect{S}) = \left\{
	\begin{array}{ll}
		1  & :  \sum_{k=1}^N S^i_{k} \geq N \\
		0  & :  \sum_{k=1}^N S^i_{k} = 0
	\end{array}
	\right.
\end{equation}
At first glance, it  is surprising that such a severe nonlinearity, applied to two MIP-correlated spiking pools, results in nearly the same performance is simple spike integration (c.f. Figure \ref{fig:MIPSI}).  The intuition here is that optimal inference requires essentially performing spike integration on the correlating spike train, as no information about the firing rate is added through spike deletions.  This random walk on one of three cases (-1,0, or +1) is approximated by linear integration, in the limit as the size of the pool ($N$) increases.

Another perspective on the nonlinearities that enable optimal computation is that they leverage knowledge about the mechanism of correlations, to improve performance.
In the SIP model, the nonlinear function depicted in Figure \ref{fig:Nonlinearity}C is, as in the MIP case, a consequence of applying a nonlinearity to each pool, and then subtracting.  However, in this case, the form is not as drastic---a shared spike event coming from the correlating train only registers as a single spike:
\begin{equation}
	W_i = f_{SIP}(\vect{S}^i_p)-f_{SIP}(\vect{S}^i_n) \label{eq:transferFunction}
\end{equation}
\begin{equation}
	f_{SIP}(\vect{S}) = \left\{
	\begin{array}{ll}
		\sum_{k=1}^N S^i_{k}  & :  \sum_{k=1}^N S^i_{k} < N \\
		1  & :  \sum_{k=1}^N S^i_{k} = N
	\end{array}
	\right.
\end{equation}
Intuitively, this strategy uses the fact the a simultaneous spike in every neuron in a pool only has one explanation for a sufficiently small window of integration, and therefore uses the correlating spike train as an additional independent input in the likelihood ratio.  At low values of $\theta$, this does not confer much of an advantage; however as the threshold increases, higher accuracy is achievable at much shorter decision times. The nonlinearity is pictured Figure \ref{fig:SIPSINonlinear}A, and also offers an intuition as to why, for low threshold values, spike integration performs almost optimally: when spikes from the correlating train are rare (or can be properly weighted), spike integration implements SPRT (Figure \ref{fig:SIPSINonlinear}B).

\section{Discussion}

\begin{figure}
	\centering
	\includegraphics[width=3in]{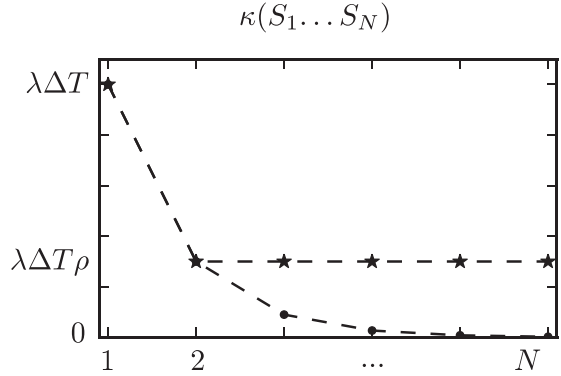}
	\caption{The joint cumulants of the SIP and MIP processes differ for pools of  greater than two neurons. Under the additive (SIP) model, the joint cumulants of the spike counts from $N$ neurons is constant for all $N>2$.  In contrast, the joint cumulants of the subtractive (MIP) model decay geometrically as the pool size increases, and this difference helps to characterized the differences in higher-order correlations between the two models. (See Appendix \ref{sec:JointCumulantAppendix} for supplementary computations.)}
	\label{fig:Cumulant}
\end{figure}


Correlated spiking among the neurons that encode sensory evidence appear ubiquitous.  Such correlations might arise arise from any number of neuroanatomical features -- the simplest being overlapping feedforward connectivity which can cause collective fluctuations across a population~\cite{Binder:2001ti,Shadlen:1998ta,DeLaRocha:2007go,Mazurek:2003cm}.  They can also result from sensory events that impact an entire population, or from rapid modulatory effects.  Moreover, for large neural populations it appears that accurate descriptions of population-wide activity can require more than the typically measured pairwise correlations, but higher order interactions as well ~\cite{Montani:2009js,Ganmor:2011ct,Yu:2011ir}.

The aim of our study is to improve our understanding of how correlated activity in these populations can impact the speed and accuracy of decisions that require accumulating sensory information over time.  Faced with the wide range of possible mechanisms and structures of correlations alluded to above, we choose to focus on two models for population-wide correlations that illustrate a key distinction in how correlations can occur.  These models have identical first-order and pairwise statistics, but differ in how each common spiking events either involves a small subset of the neurons (the subtractive, MIP case) or each neuron in the pool (the additive, SIP case) \cite{Kuhn:2003tq,springerlink:10.1007/978-1-4419-5675-0_12}.    

Figure \ref{fig:Cumulant} quantifies this difference:  based on calculations in Appendix \ref{sec:JointCumulantAppendix}, we plot the joint cumulant across $k$ neurons in a pool under both subtractive (MIP) and additive (SIP) correlations.  While the additive model possesses a constant joint cumulant no matter how many neurons are included, the joint cumulant of $k$ neurons falls off geometrically for the subtractive case.  We conjecture that this is a statistical signature that could suggest when other, more general patterns of correlated activity -- measured experimentally or arising in mechanistic models of neural circuits~\cite{Mazurek:2003cm} -- will produce similar effects on decisions. Exploring this conjecture via models and data is a target of our future research.

We summarize our main findings are as follows.  For both models of correlated spiking, decisions produced by a simple, linear spike integration model (i.e., a neural integrator) become slower and less accurate as correlations increase. 
However, a strong difference appears for decisions made via the optimal decision strategy (SPRT).  Here, additive correlations have only a minor impact on decision performance, while subtractive correlations continue to strongly diminish this performance.  The conclusion is that decision making circuits, faced with subtractive (MIP) correlated sensory populations, will invariably produce diminished decision performance, and  stand little to gain by implementing computations more complex that a simple integration of spikes over time and neurons.  However, in the presence of additive (SIP) correlations, circuit mechanisms that implement or approximate the SPRT -- perhaps via a nonlinearity such as that shown in Fig.~\ref{fig:SIPSINonlinear} applied to the sum of incoming spikes -- stand to produce substantially better decision performance than their linear counterparts.


In other contexts, nonlinear computations have also been shown to improve discrimination between two alternatives.  Field and Rieke \cite{Field:2002we,Field:2005vo} demonstrated the importance of a thresholding nonlinearity in pooling the responses of rod cells, where this nonlinearity served to reject ``background" noise.  Closer to the present setting, gating inhibition that prevents accumulation of noise samples before the onset of evidence-encoding stimulus can account for visual search performance \cite{Purcell:2012kr}, and recent results suggest that related nonlinearities can improve performance for mistuned neural integrators (\cite{Cain:ta}, see also \cite{Cain:2012ky}). 

Our cases in which correlations decrease performance -- in particular, when spikes are linearly integrated -- are consistent with several prior studies of the role of correlated activity in decision making~\cite{Zohary:1994uv,Britten:1996vr,Cohen:2009tt}.  We note, however, two differences in our models.  The first is the mechanism through which correlated spikes are generated; while we use additive and subtractive models based on Poisson processes, the authors of~\cite{Britten:1996vr,Cohen:2009tt} use a multivariate Gaussian description of spike counts. 
The second is that in \cite{Britten:1996vr,Cohen:2009tt}, decisions are rendered after a duration that is fixed before the trial begins (either a single duration,~\cite{Britten:1996vr}, or one that is drawn from a distribution of reaction times, in~\cite{Cohen:2009tt}).  This is different from the setting here, where incoming signal on each trial determines the reaction time through a bound crossing.  

Our result, in the case of subtractive (MIP) correlations, that linear integration of spikes closely approximates the optimal decision making strategy is similar to findings of Beck et al. \cite{Beck:2008db}.  Specifically, they model a dense range of differently tuned populations, and find that optimal Bayesian inference can be based on linear integration of inputs, for a wide set of correlation models.  Our additive (SIP) case, however, behaves differently, as nonlinearities are needed to achieve the optimal strategy.

An aim of future work is extending the setting of our study to include  tuning curves as in~\cite{Cohen:2009tt,Beck:2008db,Zhang:2010tj,Cohen:2009tt}.  This is more realistic for many decision tasks (including the direction discrimination task), and will also allow progress toward models with multiple decision alternatives.  An important challenge will come from defining pairwise correlations that vary as a function of preferred tuning orientation (see \cite{Zohary:1994uv,Cohen:2008js}), while also including the full structure of correlating events across multiple cells in a realistic way.  For example, in the present paper, additive correlating events occurred independently in the two populations; future work could take a more graded approach, in which some events impact the entire sensory population (i.e., as in an eyeblink or possibly an attentional shift during a visual task).  

As long as each neuron remains modeled as a Poisson point process, the sequential accumulation theory utilized here will carry over directly.  This points to another limitation of the present study and opportunity for future work.  This is the lack of temporal correlations in the statistics of the inputs.  
A model of correlations that includes spikes from a correlating train that are temporally jittered \cite{Gutnisky:2010cp, springerlink:10.1007/978-1-4419-5675-0_12} could provide a starting place for a model of the input trains, however defining updates to the likelihood ratio for the two competing hypotheses will be more difficult.  Nevertheless, it will be interesting to see how our results carry over; in particular, there will be many more different combinations of spike events that will contribute to increments for both spike integration and SPRT decision models.  

While we therefore view the present study as a first step in exploring many possibilities, our findings demonstrate how the population-wide structure of correlations -- beyond pairwise correlation coefficients -- can strongly impact the speed and accuracy of decisions, and the circuit operations necessary to achieve optimal performance.   This suggests that multi-electrode and imaging technologies, together with theoretical work on neural coding, will continue to play an exciting role in understanding the structure of basic computations like decision making over time.


\paragraph{Acknowledgements:}
We thank Yu Hu, Adrienne Fairhall, and Michael Shadlen for their valuable comments on the manuscript. We gratefully acknowledge the support of a Career Award at the Scientific Interface from the Burroughs Welcome Fund and NSF Grant CAREER DMS-1026125 (E. S. B.), and the University of Washington eScience Institute's Hyak computer cluster.

\appendix

\section{Sequential Probability Ratio Test}

\subsection{Nontrivial root of the moment generating function  (SPRT)}
\label{sec:MGF_SPRT}

The nontrivial real root of the moment generating function (MGF) of a sampling distribution is critical to finding $FC$ and $DT$ of an independently sampled sequential hypothesis test (via Equations \ref{eq:FC} and \ref{eq:RT}).  For the SPRT, the increment distribution is given in Equation \ref{eq:SPRTSamplingRV} as:
\begin{equation}
	W_i = \log\left[	
	   \frac
	   {P[\vect{S^i_p},\vect{S^i_n}|H_1]}
	   {P[\vect{S^i_p},\vect{S^i_n}|H_0]}	
	   	\right] \label{eq:SPRTSamplingRVAppendix}
\end{equation}
The ``correct" hypothesis $H_1$ is in the numerator in order to orient a crossing of the positive decision threshold with a correct choice.
Correspondingly, the probability of observing a {\it given} sample $\vect{S^i_p}$,$\vect{S^i_n}$ is known from assumption of this hypothesis, and by definition follows the distribution:
\begin{equation}
	P[\vect{S^i_p},\vect{S^i_n}|H_1]=P[\vect{S^i_p}|H_1]P[\vect{S^i_n}|H_1],
\end{equation}
where the independence assumption of the spike count vectors from the two separate pools $\vect{S^i_p}$ and $\vect{S^i_n}$ have allowed the factoring of the distribution.
Dropping the sampling index $i$ for notational convenience, the MGF can then be computed as:
\begin{equation}
	\phi_W(s)=E[e^{sW}]=\sum_{\vect{s_p},\vect{s_n}} P[\vect{s_p}|H_1]P[\vect{s_n}|H_1] \left(	   \frac{P[\vect{s_p}|H_1]P[\vect{s_n}|H_1]}{P[\vect{s_p}|H_0]P[\vect{s_n}|H_0]}		\right)^{s} 
\end{equation}
The nontrivial root ($h_0\neq0$) can then be seen by inspection (cf. Equation \ref{eq:IndSPRTH0}):
\begin{equation}
	s= h_0=-1 \;\; \implies \;\; \phi_W(h_0)=1 \label{eq:IndSPRTH0Appendix}
\end{equation}
We note that this computation is fully general, without any assumptions on the structure of correlations both within and across pools.

\subsection{$E[w]$, Independent interactions (SPRT)}

The {other} parameter of the sampling distribution critical to computing the $FC$ and $DT$ functions, $E[W]$, is computed for independent spike count distributions ($\rho=0$, cf. \ref{eq:IndSPRTEz}) as follows (see also \cite{Zhang:2010tj}):  
\begin{equation}
	E[W]=\sum_{\vect{s_p},\vect{s_n}} P[\vect{s_p}|H_1]P[\vect{s_n}|H_1] \log\left[	   \frac{P[\vect{s_p}|H_1]P[\vect{s_n}|H_1]}{P[\vect{s_p}|H_0]P[\vect{s_n}|H_0]}		\right]
\end{equation}
\begin{equation}
	=\sum_{\vect{s_p},\vect{s_n}} P[\vect{s_p}|H_1]P[\vect{s_n}|H_1] 
	\left(
		\log	 P[\vect{s_p}|H_1]+
		\log	 P[\vect{s_n}|H_1]-
		\log	 P[\vect{s_p}|H_0]-
		\log	 P[\vect{s_n}|H_0]
	\right)
\end{equation}
\begin{eqnarray}
	&=& \sum_{\vect{s_p}} P[\vect{s_p}|H_1] \log	 P[\vect{s_p}|H_1]+
  	  \sum_{\vect{s_n}} P[\vect{s_n}|H_1] \log	 P[\vect{s_n}|H_1] \\
	&&-  \sum_{\vect{s_p}} P[\vect{s_p}|H_1] \log	 P[\vect{s_p}|H_0]-
	  \sum_{\vect{s_n}} P[\vect{s_n}|H_1] \log	 P[\vect{s_n}|H_0] \label{eq:preInd}
\end{eqnarray}
\begin{equation}
	= N \left(
			E\left[\log \frac{P[S_p|H_1]}{P[S_p|H_0]}|H_1\right] +
			E\left[\log \frac{P[S_n|H_1]}{P[S_n|H_0]}|H_1\right] 
		\right) \label{eq:postInd}
\end{equation}
\begin{equation}
	= N \Delta T \left(
			\lambda_n-\lambda_p+\lambda_p \log \frac{\lambda_p}{\lambda_n} +
			\lambda_p-\lambda_n+\lambda_n \log \frac{\lambda_n}{\lambda_p} 
		\right)
\end{equation}
\begin{equation}
	N \Delta T (\lambda_p-\lambda_n)\log \frac{\lambda_p}{\lambda_n} \label{eq:poissonUncorrSPRTRT}
\end{equation}
When this quantity is substituted into Equation \ref{eq:RT}, the $\Delta T$ will cancel off, implying that $DT$ is not a function of the sampling increment size.  We compute this quantity for correlated spike count distributions next.

\subsection{$E[W]$, additively (SIP) correlated interactions (SPRT)}
\label{sec:SIPSPRTEwAppendix}
When neurons within pools are correlated, the joint PDF of the spike count vector is no longer decomposable into the product of the marginal distributions (the critical step between Equations \ref{eq:preInd} and \ref{eq:postInd}). However, an expression for $E[W]$ can be obtained in the limit as $\Delta T \rightarrow 0$, by repeatedly expanding via Taylor series about $\Delta T=0$ throughout the computation.

First, we simplify the expression for the expected increment by using the independence of the two pools:
\begin{equation}
	E[W] = E\left[\log \frac{P[\vect{S_p}|H_1]}{P[\vect{S_p}|H_0]}|H_1\right]
		 + E\left[\log \frac{P[\vect{S_n}|H_1]}{P[\vect{S_n}|H_0]}|H_1\right] \label{eq:twoPartsSIPSPRT}
\end{equation}
Next we expand each term to first order in $\Delta T$; below, we only demonstrate the expansion for the ``preferred" population; the calculation for the null pool follows by exchanging $\lambda_p$ and $\lambda_n$. In that case, by using the Law of Total Expectation conditioned on the number of spikes in the common  spike train ``shared" across the pool $\vect{\hat{S}_p}$ (which spikes at a rate $\rho \lambda_p \Delta T$), we have:
\begin{equation}
	E\left[\log \frac{P[\vect{S_p}|H_1]}{P[\vect{S_p}|H_0]}|H_1\right] 
	= E\left[E\left[\log \frac{P[\vect{S_p}|H_1]}{P[\vect{S_p}|H_0]}|\hat{S}_p,H_1\right]|H_1\right]
\end{equation}
\begin{equation}
	=\sum_{\hat{s}_p=0}^{\infty}P[\hat{S}_p=\hat{s}_p] E\left[\log \frac{P[\vect{S_p}|H_1]}{P[\vect{S_p}|H_0]}|\hat{S}_p=\hat{s}_p,H_1\right] \label{eq:lastEquivalentEqn}
\end{equation}
\begin{equation}
	=(1-\rho\lambda_p\Delta T)E\left[\log \frac{P[\vect{S_p}|H_1]}{P[\vect{S_p}|H_0]}|\hat{S}_p=0,H_1\right]+
	\Delta T\lambda_p E\left[\log \frac{P[\vect{S_p}|H_1]}{P[\vect{S_p}|H_0]}|\hat{S}_p=1,H_1\right]+
	O(\Delta T^2) \label{eq:SIPSPRTFirstExpansion}
\end{equation}
Taking the case of $\hat{S}_p=0$, 
\begin{equation}
	E\left[\log \frac{P[\vect{S_p}|H_1]}{P[\vect{S_p}|H_0]}|\hat{S}_p=0,H_1\right]
	= \sum_{\vect{s_p}} P[\vect{s_p}|\hat{S}_p=0,H_1] \log \frac{P[\vect{s_p}|H_1]}{P[\vect{s_p}|H_0]} \label{eq:SIPDeriv1}
\end{equation}

The aim here is to take advantage of the conditioning; because the spike counts of neurons within the same pool are conditionally independent, given the number of spikes in the correlating spike train, the joint distribution across the vector $\vect{s_p}$ becomes the product of the conditioned marginal distributions.  However, this is only true for the first  factor in the summand of Equation \ref{eq:SIPDeriv1}.  To continue, we must  expand the log-ratio of the probability distributions, using the law of total probability, in $\Delta T$:
\begin{eqnarray}
	P[\vect{s_p}|H_1] &=& \sum_{\hat{s}_p}^{\infty}P[\hat{S}_p|H_1]P[\vect{s_p}|\hat{S}_p=\hat{s}_p,H_1]\\
	&=& (1-\rho\lambda_p\Delta T)P[\vect{s_p}|\hat{S}_p=0,H_1] + 
	\rho\lambda_p\Delta T P[\vect{s_p}|\hat{S}_p=1,H_1] + O(\Delta T^2)
\end{eqnarray}
\begin{eqnarray}
	P[\vect{s_p}|H_0] &=& \sum_{\hat{s}_p}^{\infty}P[\hat{S}_p|H_0]P[\vect{s_p}|\hat{S}_p=\hat{s}_p,H_0]\\
	&=& (1-\rho\lambda_n\Delta T)P[\vect{s_p}|\hat{S}_p=0,H_0] + 
	\rho\lambda_n\Delta T P[\vect{s_p}|\hat{S}_p=1,H_0] + O(\Delta T^2) 
\end{eqnarray}
Moreover, the $N-$term summation in Equation \ref{eq:SIPDeriv1} need only be over $s_i \in \{0,1\}$, as higher values will produce contributions of higher than first order in $\Delta T$. Two cases emerge for the expansion: if $s_i=0$ for any $i$, $P[\vect{s_p}|\hat{S}_p=1,H_1] = P[\vect{s_p}|\hat{S}_p=1,H_0] = 0$, and we have:
\begin{equation}
	\log \frac{P[\vect{s_p}|H_1]}{P[\vect{s_p}|H_0]} =
		\log \frac{P[\vect{s_p}|\hat{S}_p=0,H_1]}{P[\vect{s_p}|\hat{S}_p=0,H_0]} \label{eq:SPRTSIPRecycle}
\end{equation}
On the other hand, if $s_i=1$ for all $i$, we can compute the expression directly via total probability, as there are only four possible ways for the event to originate; to first-order in $\Delta T$, this is:
\begin{eqnarray}
	\log \frac{P[\vect{s_p}=\vect{1}|H_1]}{P[\vect{s_p}=\vect{1}|H_0]} &=& \sum_{i=0}^1\sum_{\hat{s}_p=0}^1 P[\vect{s_p}=\vect{1}|\hat{S}_p=\hat{s}_p,H_i]P[\hat{S}_p=\hat{s}_p,H_i] \\
	&=& \rho (\lambda_p+\lambda_n) \Delta T +O(\Delta T^N)
\end{eqnarray}
Therefore, this single element of the sum offers no order one contribution (it is multiplied by $P[\vect{s_p}=\vect{1}|\hat{S}_p=0,H_1]$ which is itself is O$(\Delta T^N)$); thus,
\begin{equation}
	\sum_{\vect{s_p}} P[\vect{s_p}|\hat{S}_p=0,H_1]\log \frac{P[\vect{s_p}|\hat{S}_p=0,H_1]}{P[\vect{s_p}|\hat{S}_p=0,H_0]} = N(1-\rho)\Delta T \left( \lambda_n - \lambda_p+\lambda_p \log \frac{\lambda_p}{\lambda_n}\right) + O(\Delta T^2)\label{eq:SIPSPRTFirstPart}
\end{equation}
The case of $\hat{s}_p=1$ is simpler, as only zero-order terms  must be kept (due to the coefficient in Equation \ref{eq:SIPSPRTFirstExpansion}). Recycling the expansion from Equation \ref{eq:SPRTSIPRecycle}, we have that to zero-order:
\begin{equation}
	E\left[\log \frac{P[\vect{S_p}|H_1]}{P[\vect{S_p}|H_0]}|\hat{S}_p=1,H_1\right]
	= \sum_{\vect{s_p}} P[\vect{s_p}|\hat{S}_p=1,H_1] \log \frac{P[\vect{s_p}|H_1]}{P[\vect{s_p}|H_0]} = \log \frac{\lambda_p}{\lambda_n} + O(\Delta T) \label{eq:SPRTSIPLast}
\end{equation}
Finally, combining Equations \ref{eq:SIPSPRTFirstExpansion}, \ref{eq:SIPSPRTFirstPart}, and \ref{eq:SPRTSIPLast}, we have that:
\begin{eqnarray}
	E\left[\log \frac{P[\vect{S_p}|H_1]}{P[\vect{S_p}|H_0]}|H_1\right] &=& 
	 (1-\rho \lambda_p \Delta T)
	 \left( 
	 	N(1-\rho)\Delta T \left( \lambda_n - \lambda_p+\lambda_p \log \frac{\lambda_p}{\lambda_n}\right) + O(\Delta T^2)
	 \right) \\
	 &+& 
	 \Delta T \rho \lambda_p
	 \left(
		 \log \frac{\lambda_p}{\lambda_n} + O(\Delta T^2)
	 \right)
\end{eqnarray}
Repeating the exercise for the other component of Equation \ref{eq:twoPartsSIPSPRT} amounts to exchanging ``$p$" for ``$n$"; adding everything together gives the final result, to first-order in $\Delta T$:
\begin{equation}
	E[W] = \left(N(1-\rho)+\rho\right) \Delta T (\lambda_p-\lambda_n) \log \frac{\lambda_p} {\lambda_n}+O(\Delta T^2)
\end{equation}
We note here that as $\rho\rightarrow0$ and $\rho\rightarrow1$, we reproduce the results that would be expected from Equation \ref{eq:poissonUncorrSPRTRT}.  Also, a more intuitive and tractable computation can be done for an analogous additively-correlated Bernoulli process, resulting in the same solution.

\subsection{$E[W]$, subtractive (MIP) correlations within pools (SPRT)}
\label{sec:MIPSPRTAppendix}

In the case of subtractive correlations within pools, the derivation of $E[W]$ is the same as the additive correlation case, up to Equation \ref{eq:lastEquivalentEqn}. In this case, however, we now have:
\begin{equation}
	E[W]=(1-\frac{\lambda_p}{\rho}\Delta T)E\left[\log \frac{P[\vect{S_p}|H_1]}{P[\vect{S_p}|H_0]}|\hat{S}_p=0,H_1\right]+
	\frac{\lambda_p}{\rho} \Delta T E\left[\log \frac{P[\vect{S_p}|H_1]}{P[\vect{S_p}|H_0]}|\hat{S}_p=1,H_1\right]+
	O(\Delta T^2) \label{eq:MIPSPRTMainEq}
\end{equation}
Taking the $\hat{S}_p=0$ case first, we notice that it is impossible for any spikes to occur without a spike in the correlating spike train:
\begin{equation}
	\vect{S_p} \neq \vect{0} \;\; \implies  \;\; P[\vect{S_p}|\hat{S}_p=0,H_1]=0 
\end{equation}
Because of this, we can simplify:
\begin{eqnarray}
	E\left[\log \frac{P[\vect{S_p}|H_1]}{P[\vect{S_p}|H_0]}|\hat{S}_p=0,H_1\right] & = & P[\vect{0}|\hat{S}_p=0,H_1]\log \frac{P[\vect{0}|H_1]}{P[\vect{0}|H_0]}\\
	& = & \log \frac{P[\vect{0}|H_1]}{P[\vect{0}|H_0]} \label{eq:MIPSPRTcoeff0}
\end{eqnarray}
Interestingly, after conditioning on the number of correlating spikes, the probability of the zero vector (or any vector $\vect{s_p}$) is the same under both $H_0$ and $H_1$:
\begin{equation}
	P[\vect{0}|\hat{S}_p=\hat{s}_p,H_0] = P[\vect{0}|\hat{S}_p=\hat{s}_p,H_1] \end{equation}
We then expand to first-order in $\Delta T$:
\begin{eqnarray}
	\log \frac{P[\vect{0}|H_1]}{P[\vect{0}|H_0]} &=& 
	\log \frac{\left(1-\frac{\lambda_p}{\rho} \Delta T  \right) + \frac{\lambda_p}{\rho}\Delta T P[\vect{0}|\hat{S}_p=1]+O(\Delta T^2)}
	{\left(1-\frac{\lambda_n}{\rho} \Delta T  \right) + \frac{\lambda_n}{\rho} \Delta T P[\vect{0}|\hat{S}_p=1]+O(\Delta T^2)}\\
	&=& \frac{(\lambda_p - \lambda_n) \Delta T}{\rho}\left((1-\rho)^N-1\right) +O(\Delta T^2) \label{eq:noContribAtZeroOrder}
\end{eqnarray}
In the case of $\hat{S}_p=1$, only zero-order terms must be computed. When computing 
\begin{eqnarray}
	E\left[\log \frac{P[\vect{S_p}|H_1]}{P[\vect{S_p}|H_0]}|\hat{S}_p=1,H_1\right] = \sum_{\vect{s_p}} P[\vect{s_p}|\hat{S}_p=1,H_1]\log \frac{P[\vect{s_p}|H_1]}{P[\vect{s_p}|H_0]},
\end{eqnarray}
the summation only carries over $\{0,1\}$ for each element of $\vect{s_p}$. The case of $\vect{s_p}=\vect{0}$ provides no contribution at zero-order, as can be seen by Equation \ref{eq:noContribAtZeroOrder}; for any other case, there will be a degeneracy in the expansion of the log, caused by an absence of order 0 terms:
\begin{eqnarray}
	\vect{s_p}\neq 0 \;\; \implies \;\; \log \frac{P[\vect{s_p}|H_1]}{P[\vect{s_p}|H_0]}
	&=& \log \frac{\left(\frac{\lambda_p}{\rho} -\frac{\lambda_p^2}{\rho^2}\Delta T  \right)P[\vect{s_p}|\hat{S}_p=1] + \frac{\lambda_p^2}{2\rho^2}P[\vect{s_p}|\hat{S}_p=2]+...}
	{\left( \frac{\lambda_n}{\rho} -\frac{\lambda_n^2}{\rho^2}\Delta T \right)P[\vect{s_p}|\hat{S}_p=1]+ \frac{\lambda_n^2}{2\rho^2}P[\vect{s_p}|\hat{S}_p=2]+...}\\
	&=& \log \frac{\lambda_p}{\lambda_n} + O(\Delta T) 
\end{eqnarray}
Therefore, to first-order in $\Delta T$,
\begin{eqnarray}
	E\left[\log \frac{P[\vect{S_p}|H_1]}{P[\vect{S_p}|H_0]}|\hat{S}_p=1,H_1\right]
	&=&
	 \log \frac{\lambda_p}{\lambda_n}\sum_{\vect{s_p}\neq 0}P[\vect{s_p}|\hat{S}_p=1,H_1] + O(\Delta T)\\
	 &=&
	 \log \frac{\lambda_p}{\lambda_n}(1-(1-\rho)^N) + O(\Delta T)
	 \label{eq:lastMIPSPRTEqToUse}
\end{eqnarray}
Combining Equations \ref{eq:MIPSPRTMainEq}, \ref{eq:MIPSPRTcoeff0}, \ref{eq:noContribAtZeroOrder}, and \ref{eq:lastMIPSPRTEqToUse}, we find that:
\begin{equation}
	E\left[\log \frac{P[\vect{S_p}|H_1]}{P[\vect{S_p}|H_0]}|H_1\right] =
	\frac{
	\left(1-(1-\rho)^N \right) \left(\lambda_n-\lambda_p + \lambda_p \log \frac{\lambda_p}{\lambda_n}\right)
	}{\rho}\Delta T
\end{equation}
As before, exchanging ``$p$" for ``$n$" takes care of the expression for the null pool, and adding together gives:
\begin{equation}
	E[W] =  \frac{\left(1-(1-\rho)^N \right)}{\rho}(\lambda_p-\lambda_n)\log \frac{\lambda_p} {\lambda_n} \Delta T+O(\Delta T^2)
\end{equation}
Once again, as $\rho\rightarrow0$ and $\rho\rightarrow1$, we reproduce the results that would be expected from Equation \ref{eq:poissonUncorrSPRTRT}.

\section{Spike integration}

\subsection{Independent spiking (SI)}
\label{sec:SIuncorr}
Computing $FC$ and $DT$ for the spike integration accumulation model relies on computation of the MGF for the sampling distribution.  We begin with several identities that will be useful below. 
The MGF for the sum of N independent random variables is:
\begin{equation}
	S = \sum_{i=1}^N S_i 
	\;\; \iff \;\;
	\phi_S(t) = \phi_{s_i}(t)^N
\end{equation}
Given that the MGF for a random variable $S=\phi_S(t)$, it follows that
\begin{equation}
	\phi_{-S}(t) = \phi_{S}(-t)
\end{equation}
Finally, the MGF for a Poisson random variable is:
\begin{equation}
	\phi(t)=e^{\lambda(e^t-1)}
\end{equation}
Given the definition of the increment variable in Equation \ref{eq:SISamplingRV}, and noting that each spike count random variable is independent, we can combine these observations to construct the MGF for the sampling random variable, over a time window $\Delta T$:
\begin{equation}
	\phi_W(t)=(e^{\lambda_p\Delta T(e^t-1)})^N (e^{\lambda_n \Delta T(e^{-t}-1)})^N
\end{equation}
Now the nontrivial root can be calculated (cf. Equation \ref{eq:uncorrSI}):
\begin{equation}
	h_0 = -\log \left( \frac{\lambda_p}{\lambda_n}\right) \;\;
	\implies \;\;
	 \phi_W(h_0)=1
\end{equation}
Because the MGF is known explicitly, the computation of the expected increment is simple (cf. Equation \ref{eq:uncorrSIEz}):
\begin{equation}
	E[W]=\phi_W'(0)=\Delta T N\left( \lambda_p-\lambda_n \right).\label{eq:uncorrSIEzAppendix}
\end{equation}

\subsection{Additive (SIP) correlated interactions within pools (SI)}

When additive correlations are introduced within pools, the spike count distribution MGF over a time period $\Delta T$ can still be broken into the product of two separate MGF's, one each for the preferred and null pools, which are identical in form but differ in their Poisson rate parameters (indicated by the semicolon):
\begin{equation}
	\phi_W(t) = \phi_{S}(t;\lambda_p \Delta T) \phi_{S}(-t;\lambda_n \Delta T) \label{eq:uncorrPoolSIRelation}
\end{equation}
For the preferred pool, the spike count can be broken into two independent contributions---spikes $\hat{S}$ from the shared (i.e. correlating) spike train that get counted $N$ times (firing at a rate $\rho \lambda$), and spikes from the $N$ independent spike trains that get counted once (each firing at a rate $(1-\rho) \lambda$):
\begin{equation}
	\phi_{S}(t;\lambda) = \phi_{\hat{S}}(t;\lambda)\phi_{S_i}(t;\lambda)^N
\end{equation}
The MGF for the shared spike train can be computed directly from the definition, using its probability mass function (PMF):
\begin{equation}
   P[\hat{S}=iN] = \left\{
     \begin{array}{ll}
       \frac{e^{-\lambda}\lambda^i}{i!}  &   i \in \naturalsWithZero \\
       0 & \;\;\;\;\;\; \text{otherwise,}
     \end{array}
   \right.
\end{equation}
and thus:
\begin{equation}
	\phi_{\hat{S}}(t,\lambda)
	\;=\;
	\sum_{k=0}^\infty e^{tk}P[\hat{S}=k] 
	\;=\; 
	\sum_{k=0}^\infty e^{Ntk}\frac{e^{-\lambda}\lambda^k}{k!}
	\;=\;
	e^{(e^{Nt}-1)\lambda}
\end{equation}
The MGF for the independent spike trains $\phi_{S_i}(t,\lambda)$ follows from Section \ref{sec:SIuncorr}, giving the form of the MGF of the increment over a time $\Delta T$ as: 
\begin{equation}
	\phi_W(t) = 
	e^{(e^{Nt}-1)\rho\lambda_p \Delta T}
	e^{(e^{-Nt}-1)\rho\lambda_n \Delta T}
	\left(e^{(e^t-1)(1-\rho) \lambda_p \Delta T}\right)^N
	\left(e^{(e^{-t}-1)(1-\rho) \lambda_n \Delta T}\right)^N
\end{equation}
After rearranging, $h_0$ is implicitly defined as the nontrivial root of:
\begin{equation}
	\lambda_p(
	\rho(e^{Nt}-1)+(1-\rho)N(e^t-1)
			 ) 
	+
	\lambda_n(
	\rho(e^{-Nt}-1)+(1-\rho)N(e^{-t}-1)
			 ) = 0 \; \implies \; h_0=t
\end{equation}
As $\rho\rightarrow0$, we recover the solution from Section \ref{sec:SIuncorr}. The expected increment can be directly computed as:
\begin{equation}
	E[W]=\Delta T N\left( \lambda_p-\lambda_n \right) \\
\end{equation}
Note that this last expression is the same as in the independent case (Equation \ref{eq:uncorrSIEz}), as expected, and that unlike the SPRT, no limits in $\Delta T$ were necessary to compute the parameters  for the $FC$ and $DT$ functions.

\subsection{Subtractive (MIP) correlated interactions within pools (SI)}
\label{sec:MIPSIAppendix}

With subtractive correlations, we again derive an MGF for the spike count vector of an individual pool $\phi_{\hat{S}}(t;\lambda)$, and apply Equation \ref{eq:uncorrPoolSIRelation}.  In this case, however, the number of spikes in a pool, conditioned on the number of spikes in that pools correlating train, is binomially distributed. Thus applying the Law of Total Probability:
\begin{equation}
	P[S=s] = \sum_{\hat{s}=0}^\infty \text{Poiss}[\hat{s};\frac{\lambda}{\rho}]\text{Binom}[\hat{s}N,s;\rho]
\end{equation}
using the definitions for the PMF's of the Poiss$[i;\lambda]$ and Binom$[N,k;p]$ distributions, we have:
\begin{equation}
	\phi_{S}(t;\lambda) 
	= E[e^{St}]
	= \sum_{s=0}^\infty P[S=s]e^{st} 
	= e^{\left( 
			   ((1+\rho(e^t-1))^N -1)
		\right) \frac{\lambda}{\rho}}
\end{equation}
After applying Equation \ref{eq:uncorrPoolSIRelation} with this MGF for both the preferred and null population, we find an implicit relationship for the non-trivial real root $t=h_0$ that does not depend on $\Delta T$:
\begin{equation}
	\left( 
			   ((1+\rho(e^t-1))^N -1)
	  \right)\lambda_p + 
	\left( 
			   ((1+\rho(e^{-t}-1))^N -1)
	  \right)\lambda_n=0 \; \implies \; h_0=t
\end{equation}
As before, the expected increment can be directly computed by differentiation, and we find the same expression as in the additive correlation case:
\begin{equation}
	E[W]=\Delta T N\left( \lambda_p-\lambda_n \right) \\
\end{equation}

\section{Speed and accuracy functions with overshoot}

\label{sec:overshootAppendix}

The identities provided in Equations \ref{eq:FC} and \ref{eq:RT} are very useful, however are simplifications of the full formulas for $FC$ and $DT$ (assuming $E[W]\neq0$) derived by Wald \cite{Wald:1944th}, which are:
\begin{eqnarray}
	FC &=& 1-\frac{E[e^{h_0E_n}|E_n \geq \theta]-1}{E[e^{h_0E_n}|E_n\geq \theta]-E[e^{h_0E_n}|E_n\leq -\theta]} \label{eq:FCFull} \\
	DT &=& \frac{\Delta T}{E[W]}\left(
	E[E_n|E_n\geq \theta](FC)
	+E[E_n|E_n\leq -\theta](1-FC)\right) \label{eq:RTFull}
\end{eqnarray}
Specifically, Equations \ref{eq:FC} and \ref{eq:RT} hold under the assumption that the value of the state variable on the decision step  is exactly equal to the decision threshold.  In practice, however, this ``no-overshoot" assumption may not provide a particularly good approximation. 

A correction term based on the mean of the overshoot distribution -- that is, the distribution of the random variable defined by the excess distance over either the positive or negative threshold on the threshold crossing step --  is suggested by Lee et al \cite{Lee:1994um}. This correction is based on the Taylor expansion of the conditional expectations in Equation \ref{eq:FCFull}, and takes the form of a shift in the decision threshold. A correction of this form is relevant to our analysis, as the performance of two models are compared parametrically in the threshold to isolate the effects of the speed-accuracy tradeoff imparted by freely adjusting the threshold.

Denote the value of $E_n$ conditioned on crossing the first threshold as $\hat{E}_n$, and let $X=\hat{E}_n-\theta$ overshoot random variable, with mean $u_X$. Expanding the conditional expectation (although dropping the conditional notation for convenience) via a Taylor series centered on this mean (the so-called delta method), we have
\begin{equation}
	E[e^{h_0\hat{E}_n}] = E[e^{h_0r_0}+h_0e^{h_0r_0}(\hat{E}_n-r_0)+\frac{h_0^2e^{h_0r_0}(\hat{E}_n-r_0)^2}{2}+...]
\end{equation}
Choosing $r_0=\theta$ yields an expression of Wald's truncation:
\begin{eqnarray}
	E[e^{h_0\hat{E}_n}] &=& 
	e^{h_0\theta} \left(1
		+ h_0E[X] 
		+ \frac{h_0^2E[X^2]}{2}+... \right)\\
	& \approx & e^{h_0\theta}
\end{eqnarray}
Here we see that if $E_n=\theta$, each term in the expansion becomes zero and Wald's approximation holds exactly.    On the other hand, if $E_n$ overshoots $\theta$, error will accumulate at each term in the expansion, as a function of the moments of the overshoot distribution.  If instead the expansion is performed about $r_0=\theta+\mu_x$, a threshold-shifted approximation expresses the truncation error terms of the second and higher centered moments of the overshoot distribution: 
\begin{eqnarray}
	E[e^{h_0E_n}] & = & 
	e^{h_0(\theta+\mu_X)} \left(1
		+ \frac{h_0^2E[(X-\mu_X)^2]}{2}+...\right)\\
	& \approx & e^{h_0(\theta+E[X_n])}
\end{eqnarray}
In practice, the overshoot distribution is often nonzero; however, if its mean can be calculated and $h_0<0$, the truncation error associated with the latter approximation might provide a more favorable approximation as long as the higher-order moments do not grow too large.  For the decision time, using this alternative approximation is exactly correct, and results in no additional error. 

\section{Joint cumulants for the SIP and MIP model}
\label{sec:JointCumulantAppendix}

Staude et al. \cite{springerlink:10.1007/978-1-4419-5675-0_12} suggest that cumulants  provide a ``natural and intuitive higher-order generalization of the covariance" for multineuron spiking. The two models of correlated activity examined here are indistinguishable when only examining first-order (i.e., mean firing rate) or second-order (i.e., pairwise correlations) statistics. Here, we derive the joint cumulants for each of these two models, to clarify how the spike count distributions produced by the two models differ at higher orders.

The derivation relies on the conditional independence of the spike counts for each neuron in a pool, conditioned upon the spike count in the common spike train. Let $S_1...S_N$ be the random variables giving spike counts in a windows of size $\Delta T$ from each of the $1...N$ neurons in a correlated pool, and $\hat{S}$ be the spike count in the common spike train.  The law of total cumulance \cite{Brillinger:1969ub}  allows a relatively simple expression of the joint cumulant on $k$ members  $S_1...S_N$ (Because of the homogeneity of the pool, we will express the $k^{th}$ joint cumulant as calculated on $S_1...S_k$, but the same expression holds for any $k$-sized subset of $S_1...S_N$):
\begin{equation}
	\kappa(S_1,...,S_k)=\sum_{\pi \in \Pi} 
	\kappa 
	\left( 
		\kappa (S_{B_1}|\hat{S})  ,...,  \kappa(S_{B_b}|\hat{S})
	\right) \label{eq:condCumulantFormula}
\end{equation}
Here $\Pi$ is the set of all partitions of $\{1...k\}$, for example 
\begin{eqnarray}
\Pi[\{1,2,3\}] &=& 
	\{ 
		\{ 		\{1\},\{2\},\{3\}			\},
		\{ 		\{1,2\},\{3\}				\},
		\{		\{1\},\{2,3\}				\},
		\{ 		\{1,3\},\{2\}				\},
		\{		1,2,3						\}
	\}\\
	&=& \{ \pi_1, \pi_2,\pi_3,\pi_4,\pi_5  \} \label{eq:PartitionExample}
\end{eqnarray}
and $\kappa(S_{B_j}|\hat{S})$ is the conditional joint cumulant over the set of all spike counts indexed by an element of $B_j$---that is, the set $\{S_j:j\in B_j , B_j \in \pi_i \}$. 

In our special case, $\kappa(S_{B_j}|\hat{S}) = 0$ whenever $|B_j|>1$, owing to the conditional independence of each neuron given the common spike train.  Moreover, from the definition of the  cumulant, the term of \eqref{eq:condCumulantFormula} for the partition $\pi_i$ that contains such a block $B_j$ will also be zero. This implies that the only $\pi_i \in \Pi$ that contributes in Equation \ref{eq:condCumulantFormula} is $\pi_i=\{  \{1\} ... \{k\} \}$ ($i=1$ in the example of Equation \ref{eq:PartitionExample}); thus
\begin{equation}
	\kappa(S_1,...,S_k)=
	\kappa(E[S_1|\hat{S}],...,E[S_k|\hat{S}]) = \kappa_k(E[S_1|\hat{S}]),
\end{equation}
where we have used the fact that the first cumulant is simply the expected value.
Using the cumulant generating function, we then have a formula for the joint cumulant:
\begin{equation}
	\kappa(S_1,...,S_k) = \left. \frac{d^k}{(dt)^k} \left[\log E[e^{t E[S_1|\hat{S}] }]\right]\right|_{t=0}
\end{equation}
Thus, for the two models of correlations (assuming a firing rate $\lambda$), we have:
\begin{eqnarray}
	MIP:\;\;E[S_1|\hat{S}=\hat{s}] &=& \sum_{s_1=0}^{\hat{s}} s_1 {\hat{s} \choose s_1} \rho^{s_1} (1-\rho)^{\hat{s}-s_1} = \rho \hat{s}\\
	&\implies&	\log E[e^{t E[S_1|\hat{S}=\hat{s}] }] = \log
		\sum_{\hat{s}=0}^{\infty}		\frac{e^{-\Delta T \lambda/\rho}(\Delta T \lambda/\rho)^{\hat{s}}}{\hat{s}!}	e^{t \rho \hat{s}}\\
		&&\hspace{.1 in} = \frac{\lambda \Delta T (e^{\rho t}-1)}{\rho}\\
		&\implies& \kappa(S_1...S_k) =\left. \frac{d^k}{(dt)^k}\left[ \frac{\lambda \Delta T (e^{\rho t}-1)}{\rho}\right]\right|_{t=0} 
\end{eqnarray}
\begin{equation}
	=\boxed{\Delta T \lambda \rho^{k-1} \label{eq:MIPCumulant}}
\end{equation}
\begin{eqnarray}	
	SIP:\;\;E[S_1|\hat{S}=\hat{s}] &=& \sum_{s_1=\hat{s}}^{\infty} s_1 
	\frac{e^{-(1-\rho)\Delta T \lambda} ((1-\rho)\Delta T \lambda)^{s_1-\hat{s}} }{(s_1-\hat{s})!} = \hat{s} + \lambda \Delta T (1-\rho)\\
	&\implies&	\log E[e^{t E[S_1|\hat{S}=\hat{s}] }] = \log
		\sum_{\hat{s}=0}^{\infty}		\frac{e^{-\Delta T \lambda \rho}(\Delta T \lambda \rho)^{\hat{s}}}{\hat{s}!}	e^{t (\hat{s} + \lambda \Delta T (1-\rho))}\\
		&&\hspace{.1 in} = \lambda \Delta T (\rho[e^t-t-1]+t) \label{e.esb} \\
		&\implies& \kappa(S_1...S_k) =\left. \frac{d^k}{(dt)^k}\left[ \lambda \Delta T (\rho[e^t-t-1]+t)\right]\right|_{t=0} 
\end{eqnarray}
\begin{equation}
	=\boxed{	 \left\{
							\begin{array}{ll}
								\lambda \Delta T  & :  k=1 \\
								\lambda \Delta T \rho & :  k>1
							\end{array}
						\right. \label{eq:SIPCumulant}}
\end{equation}
Comparing Equations \ref{eq:MIPCumulant} and \ref{eq:SIPCumulant} (see also Figure \ref{fig:Cumulant}), we see agreement for $k\leq2$ as expected; these correspond the the intended firing rate and pairwise covariance of neurons within the pool.  However, for $k>2$, we see the signature of the differences in the structure of the correlations. For the MIP model, the joint cumulant decays geometrically as more and more neurons are considered. In contrast, the joint cumulant remains constant for the SIP model.


\begin{thebibliography}{10}

\bibitem{Aertsen:1989tg}
AM~Aertsen, GL~Gerstein, MK~Habib, and G.~Palm.
\newblock {Dynamics of neuronal firing correlation: modulation of" effective
  connectivity"}.
\newblock {\em Journal of Neurophysiology}, 61(5):900--917, 1989.

\bibitem{Averbeck:2006ew}
Bruno~B Averbeck, Peter~E Latham, and Alexandre Pouget.
\newblock {Neural correlations, population coding and computation}.
\newblock {\em Nature Reviews Neuroscience}, 7(5):358--366, May 2006.

\bibitem{Bair:2001wm}
Wyeth Bair, Ehud Zohary, and William~T Newsome.
\newblock {Correlated firing in macaque visual area MT: time scales and
  relationship to behavior}.
\newblock {\em Journal of Neuroscience}, 21(5):1676--1697, 2001.

\bibitem{Beck:2008db}
Jeffrey~M Beck, Wei~Ji Ma, Roozbeh Kiani, Tim Hanks, Anne~K Churchland, Jamie
  Roitman, Michael~N Shadlen, Peter~E Latham, and Alexandre Pouget.
\newblock {Probabilistic Population Codes for Bayesian Decision Making}.
\newblock {\em Neuron}, 60(6):1142--1152, January 2008.

\bibitem{Binder:2001ti}
M.D. Binder and R.K. Powers.
\newblock {Relationship between simulated common synaptic input and discharge
  synchrony in cat spinal motoneurons}.
\newblock {\em Journal of Neurophysiology}, 86(5):2266--2275, 2001.

\bibitem{Bogacz:2006fj}
Rafal Bogacz, Eric Brown, Philip Holmes, and Jonathan~D Cohen.
\newblock {The physics of optimal decision making: A formal analysis of models
  of performance in two-alternative forced-choice tasks.}
\newblock {\em Psychological Review}, 113(4):700--765, 2006.

\bibitem{Brillinger:1969ub}
David~R. Brillinger.
\newblock {The calculation of cumulants via conditioning}.
\newblock {\em Ann Inst Stat Math Annals of the Institute of Statistical
  Mathematics}, 21(1):215--218, 1969.

\bibitem{Britten:1996vr}
K~H Britten, W~T Newsome, M~N Shadlen, S.~Celebrini, and J~A Movshon.
\newblock {A relationship between behavioral choice and the visual responses of
  neurons in macaque MT}.
\newblock {\em Visual Neuroscience}, 13:87--100, 1996.

\bibitem{Britten:1992wx}
Kenneth~H Britten, Michael~N Shadlen, William~T Newsome, and J~Anthony Movshon.
\newblock {The analysis of visual motion: a comparison of neuronal and
  psychophysical performance}.
\newblock {\em The Journal of Neuroscience}, 12(12):4745--4765, 1992.

\bibitem{Britten:1993wv}
Kenneth~H Britten, Michael~N Shadlen, William~T Newsome, and J~Anthony Movshon.
\newblock {Responses of neurons in macaque MT to stochastic motion signals}.
\newblock {\em Visual Neuroscience}, 10(06):1157--1169, 1993.

\bibitem{Cain:ta}
Nicholas Cain, Andrea Barreiro, Mike Shadlen, and Eric Shea-Brown.
\newblock {A favorable tradeoff between robustness and performance in
  sequential decision tasks}.
\newblock In {\em COSYNE: 2011}, Salt Lake City, February 2011.

\bibitem{Cain:2012ky}
Nicholas Cain and Eric Shea-Brown.
\newblock {Computational models of decision making: integration, stability, and
  noise}.
\newblock {\em Current Opinion In Neurobiology}, pages 1--7, May 2012.

\bibitem{Cohen:2011eh}
Marlene~R Cohen and Adam Kohn.
\newblock {Measuring and interpreting neuronal correlations}.
\newblock {\em Nature Publishing Group}, 14(7):811--819, June 2011.

\bibitem{Cohen:2008js}
Marlene~R Cohen and William~T Newsome.
\newblock {Context-Dependent Changes in Functional Circuitry in Visual Area
  MT}.
\newblock {\em Neuron}, 60(1):162--173, October 2008.

\bibitem{Cohen:2009tt}
Marlene~R Cohen and William~T Newsome.
\newblock {Estimates of the contribution of single neurons to perception depend
  on timescale and noise correlation}.
\newblock {\em Journal of Neuroscience}, 29(20):6635--6648, 2009.

\bibitem{DeLaRocha:2007go}
Jaime De~La~Rocha, Brent Doiron, Eric Shea-Brown, Kre{\v s}imir Josi{\'c}, and
  Alex Reyes.
\newblock {Correlation between neural spike trains increases with firing rate}.
\newblock {\em Nature}, 448(7155):802--806, August 2007.

\bibitem{Field:2002we}
GD~Field and F~Rieke.
\newblock {Nonlinear signal transfer from mouse rods to bipolar cells and
  implications for visual sensitivity}.
\newblock {\em Neuron}, 34(5):773--785, 2002.

\bibitem{Field:2005vo}
GD~Field, AP~Sampath, and F~Rieke.
\newblock {Retinal processing near absolute threshold: from behavior to
  mechanism}.
\newblock {\em Annual Review of Physiology}, 67:491--514, 2005.

\bibitem{Ganmor:2011ct}
E~Ganmor, R~Segev, and E~Schneidman.
\newblock {Sparse low-order interaction network underlies a highly correlated
  and learnable neural population code}.
\newblock {\em Proceedings of the National Academy of Sciences}, 108(23):9679,
  2011.

\bibitem{Ghosh:1991wm}
B.~K. Ghosh and Pranab~Kumar Sen.
\newblock {\em {Handbook of sequential analysis}}.
\newblock M. Dekker, New York, 1991.

\bibitem{Gold:2002th}
Joshua~I Gold and Michael~N Shadlen.
\newblock {Banburismus and the Brain Decoding the Relationship between Sensory
  Stimuli, Decisions, and Reward}.
\newblock {\em Neuron}, 36(2):299--308, 2002.

\bibitem{GoldmanReview}
MS~Goldman, A~Compte, and X.J. Wang.
\newblock {Neural integrator models}.
\newblock {\em In: Squire LR (ed.) Encyclopedia of Neuroscience}, 6:165--178,
  2009.

\bibitem{Gutnisky:2010cp}
D~A Gutnisky and K~Josic.
\newblock {Generation of Spatiotemporally Correlated Spike Trains and Local
  Field Potentials Using a Multivariate Autoregressive Process}.
\newblock {\em Journal of Neurophysiology}, 103(5):2912--2930, May 2010.

\bibitem{Huang:2009gc}
X~Huang and S~G Lisberger.
\newblock {Noise Correlations in Cortical Area MT and Their Potential Impact on
  Trial-by-Trial Variation in the Direction and Speed of Smooth-Pursuit Eye
  Movements}.
\newblock {\em Journal of Neurophysiology}, 101(6):3012--3030, May 2009.

\bibitem{Kuhn:2003tq}
A~Kuhn, A~Aertsen, and S~Rotter.
\newblock {Higher-order statistics of input ensembles and the response of
  simple model neurons}.
\newblock {\em Neural Computation}, 15(1):67--101, 2003.

\bibitem{2011arXiv1109.6524L}
Peter~E Latham and Yasser Roudi.
\newblock {Role of correlations in population coding}.
\newblock {\em Arxiv preprint arXiv:1109.6524}, q-bio.NC, 2011.

\bibitem{Lee:1994um}
J.~Lee, C.~Park, and B.~Kim.
\newblock {An Estimation Method for the Excess over the Boundaries in the SPRT
  and Its Applications.}
\newblock {\em Sequential Analysis}, 13(2):127--144, 1994.

\bibitem{Mazurek:2003cm}
Mark~E Mazurek, Jamie~D Roitman, Jochen Ditterich, and Michael~N Shadlen.
\newblock {A Role for Neural Integrators in Perceptual Decision Making}.
\newblock {\em Cerebral Cortex}, 13(11):1257--1269, November 2003.

\bibitem{Montani:2009js}
F~Montani, R~A~A Ince, R~Senatore, E~Arabzadeh, M~E Diamond, and S~Panzeri.
\newblock {The impact of high-order interactions on the rate of synchronous
  discharge and information transmission in somatosensory cortex}.
\newblock {\em Philosophical Transactions of the Royal Society A: Mathematical,
  Physical and Engineering Sciences}, 367(1901):3297--3310, July 2009.

\bibitem{Newsome:1989ul}
William~T Newsome, Kenneth~H Britten, J~Anthony Movshon, and Michael~N Shadlen.
\newblock {Single neurons and the perception of motion}.
\newblock In Dominic Man-Kit Lam and Charles~D. Gilbert, editors, {\em Neural
  mechanisms of visual perception}, pages 171--198. Portfolio Pub. Co.,
  Woodlands, Tex., April 1989.

\bibitem{Niebur:2007to}
E.~Niebur.
\newblock {Generation of synthetic spike trains with defined pairwise
  correlations}.
\newblock {\em Neural Computation}, 19(7):1720--1738, 2007.

\bibitem{Purcell:2012kr}
B~A Purcell, J~D Schall, G~D Logan, and T~J Palmeri.
\newblock {From Salience to Saccades: Multiple-Alternative Gated Stochastic
  Accumulator Model of Visual Search}.
\newblock {\em Journal of Neuroscience}, 32(10):3433--3446, March 2012.

\bibitem{Ratcliff:1978wz}
Roger Ratcliff.
\newblock {A Theory of Memory Retrieval}.
\newblock {\em Psychological Review}, 85(2):59--108, 1978.

\bibitem{Salinas:2001ul}
E~Salinas and TJ~Sejnowski.
\newblock {Correlated neuronal activity and the flow of neural information}.
\newblock {\em Nature Reviews Neuroscience}, 2(8):539--550, 2001.

\bibitem{Salzman:1992wga}
C~D Salzman, C~M Murasugi, K~H Britten, and W~T Newsome.
\newblock {Microstimulation in visual area MT: effects on direction
  discrimination performance.}
\newblock {\em The Journal of Neuroscience}, 12(6):2331--2355, June 1992.

\bibitem{Shadlen:1998ta}
MN~Shadlen and WT~Newsome.
\newblock {The variable discharge of cortical neurons: implications for
  connectivity, computation, and information coding}.
\newblock {\em Journal of Neuroscience}, 18(10):3870--3896, 1998.

\bibitem{Smith:2008gv}
M~A Smith and A~Kohn.
\newblock {Spatial and Temporal Scales of Neuronal Correlation in Primary
  Visual Cortex}.
\newblock {\em Journal of Neuroscience}, 28(48):12591--12603, November 2008.

\bibitem{Softky:1993uj}
WR~Softky and C~Koch.
\newblock {The highly irregular firing of cortical cells is inconsistent with
  temporal integration of random EPSPs}.
\newblock {\em Journal of Neuroscience}, 13(1):334--350, 1993.

\bibitem{springerlink:10.1007/978-1-4419-5675-0_12}
Benjamin Staude, Sonja Gr{\"u}n, and Stefan Rotter.
\newblock {Higher-Order Correlations and Cumulants}.
\newblock In Sonja Gr{\"u}n and Stefan Rotter, editors, {\em Analysis of
  Parallel Spike Trains}, pages 253--280. Springer US, 2010.

\bibitem{Tuckwell:1989us}
Henry~C. Tuckwell.
\newblock {\em {Stochastic processes in the neurosciences}}.
\newblock Society for Industrial and Applied Mathematics, Philadelphia, Pa.,
  1989.

\bibitem{Wald:1944th}
A~Wald.
\newblock {On cumulative sums of random variables}.
\newblock {\em Annals Of Mathematical Statistics}, 15:342--342, 1944.

\bibitem{Wald:1948uha}
A~Wald and J~Wolfowitz.
\newblock {Optimum character of the sequential probability ratio test}.
\newblock {\em The Annals of Mathematical Statistics}, 19(3):326--339, 1948.

\bibitem{Wang:2002um}
Xiao-Jing Wang.
\newblock {Probabilistic decision making by slow reverberation in cortical
  circuits}.
\newblock {\em Neuron}, 36(5):955--968, 2002.

\bibitem{Yu:2011ir}
S.~Yu, H.~Yang, H.~Nakahara, G.S. Santos, D.~Nikoli{\'c}, and D.~Plenz.
\newblock {Higher-order interactions characterized in cortical activity}.
\newblock {\em The Journal of Neuroscience}, 31(48):17514--17526, 2011.

\bibitem{Zhang:2010tj}
Jiaxiang Zhang and Rafal Bogacz.
\newblock {Optimal Decision Making on the Basis of Evidence Represented in
  Spike Trains}.
\newblock {\em Neural Computation}, 22(5):1113--1148, 2010.

\bibitem{Zohary:1994uv}
Ehud Zohary, Michael~N Shadlen, and William~T Newsome.
\newblock {Correlated neuronal discharge rate and its implications for
  psychophysical performance}.
\newblock {\em Nature}, 370:140--143, 1994.

\end{thebibliography}

\end{document}